\documentclass[12pt,twoside]{article}
\usepackage{amssymb,amsmath,graphicx,amsmath}
\usepackage[colorlinks=true,linkcolor=blue,urlcolor=blue,citecolor=blue]{hyperref}

\usepackage{times}
\usepackage{color}

\graphicspath{{./}{}}
\newcommand{\uu}{{u}}

\usepackage{pifont}

\usepackage{bbm}
\skip\footins 39pt plus 4pt minus 2pt
\def\footnoterule{\kern-19pt\hrule width.5in\kern18.6pt}\floatsep 12pt plus 2pt minus 2pt
\renewcommand{\baselinestretch}{1}
\renewcommand{\epsilon}{\varepsilon}

\graphicspath{{}}

\begin{document}

\newcommand{\red}{\color{red}}

\newcommand{\blue}{\color{blue}}
\newcommand{\fehler}[1]{{\color{red}#1}}
\newcommand{\qed}{\hfill\raisebox{-0.5mm}[0mm][0mm]{$\Box$}}
\newcommand{\st}{\!\!\!/}
\newcommand{\ksl}{k\!\!\!/}
\newcommand{\dsl}{\partial\!\!\!/}
\newcommand{\Asl}{A\!\!\!/}
\newcommand{\Dsl}{D\!\!\!\!/}
\newcommand{\half}{\frac{1}{2}}
\newcommand{\tbt}{{\bar \theta}\theta}
\newcommand{\A}{\alpha}
\newcommand{\B}{\beta}
\newcommand{\G}{\gamma}
\newcommand{\D}{\delta}
\newcommand{\E}{\varepsilon}
\newcommand{\T}{\theta}
\newcommand{\ts}{
   {\raisebox{-0.5ex}{\parbox{0.5ex}{
      \setlength{\unitlength}{0.5ex}
      \begin{picture}(1,6)
         \thinlines
         \put(0.5,-1){\line(0,1){7}}
      \end{picture}
   }}}
}
\newcommand{\lts}{
   {\raisebox{0ex}{\parbox{0.5ex}{
      \setlength{\unitlength}{0.5ex}
      \begin{picture}(1,12)
         \thinlines
         \put(0.5,-0.25){\line(0,1){12}}
      \end{picture}
   }}}
}
\newcommand{\N}{\mathbb{N}}
\newcommand{\R}{\mathbb{R}}
\newcommand{\Rscript}{\scriptstyle\mbox{\scriptsize \rm I}\!\mbox{\scriptsize\rm R}}
\newcommand{\rme}{{\mathrm{e}}}
\newcommand{\rmd}{{\mathrm{d}}}
\newcommand{\tr}{\mbox{tr}}
\newcommand{\nn}{\nonumber}
\newcommand{\weiter}{\nonumber \\ & & }
\newcommand{\nicht}[1]{}
\newlength{\lang}
\newcommand {\eq}[1]{(\ref{#1})}
\newcommand {\Eq}[1]{Eq.\hspace{0.55ex}(\ref{#1})}
\newcommand {\Eqs}[1]{Eqs.\hspace{0.55ex}(\ref{#1})}
\newcommand {\Sec}[1]{section~\ref{#1}}
\newcommand {\ds}{\displaystyle}
\newcommand {\tx}{\textstyle}
\newcommand {\scr}{\scriptstyle}
\newcommand {\scrscr}{\scripscriptstyle}
\newcommand {\ind}[1]{\mathrm{#1}}
\newcommand {\p}{\partial}
\newcommand{\trint}{\int\!\!\!\int\!\!\!\int}
\newcommand{\bn}{:\hspace*{-0.5ex}}
\newcommand{\en}{\hspace*{-0.5ex}:}
\newcommand{\1}{\,{\bf 1}\,}
\newcommand{\llongrightarrow}{-\!\!\!-\!\!\!\!\rightarrow}
\newcommand{\lllongrightarrow}{-\!\!\!-\!\!\!-\!\!\!\!\rightarrow}
\newcommand{\llllongrightarrow}{-\!\!\!-\!\!\!-\!\!\!-\!\!\!\!\rightarrow}
\newcommand{\lllllongrightarrow}{-\!\!\!-\!\!\!-\!\!\!-\!\!\!-\!\!\!\!\rightarrow}
\newcommand{\llllllongrightarrow}{-\!\!\!-\!\!\!-\!\!\!-\!\!\!-\!\!\!-\!\!\!\!\rightarrow}
\newcommand{\lllllllongrightarrow}{-\!\!\!-\!\!\!-\!\!\!-\!\!\!-\!\!\!-\!\!\!-\!\!\!\!\rightarrow}
\newcommand{\llllllllongrightarrow}{-\!\!\!-\!\!\!-\!\!\!-\!\!\!-\!\!\!-\!\!\!-\!\!\!-\!\!\!\!\rightarrow}
\newcommand{\lllllllllongrightarrow}{-\!\!\!-\!\!\!-\!\!\!-\!\!\!-\!\!\!-\!\!\!-\!\!\!-\!\!\!-\!\!\!\!\rightarrow}
\newcommand{\llllllllllongrightarrow}{-\!\!\!-\!\!\!-\!\!\!-\!\!\!-\!\!\!-\!\!\!-\!\!\!-\!\!\!-\!\!\!-\!\!\!\!\rightarrow}
\newcommand{\lllllllllllongrightarrow}{-\!\!\!-\!\!\!-\!\!\!-\!\!\!-\!\!\!-\!\!\!-\!\!\!-\!\!\!-\!\!\!-\!\!\!-\!\!\!\!\rightarrow}
\newcommand{\llllllllllllongrightarrow}{-\!\!\!-\!\!\!-\!\!\!-\!\!\!-\!\!\!-\!\!\!-\!\!\!-\!\!\!-\!\!\!-\!\!\!-\!\!\!-\!\!\!\!\rightarrow}
\newcommand{\setcurrentlabel}[1]{\def\@currentlabel{#1}}
\newbox{\expbox}
\newlength{\explength}
\newcommand{\EXPhelp}[6]{\sbox{\expbox}{\ensuremath{#4#1}}\settowidth{\explength}{\rotatebox{90}{\ensuremath{#4\left#5\usebox{\expbox}\right#6}}}\ensuremath{#4\left#5\usebox{\expbox}\right#6_{{#4\hspace*{0.4ex}}\hspace*{-0.22\explength}#2}^{{#4\hspace*{0.4ex}}\hspace*{-0.22\explength}#3}}}
\newcommand{\EXPdu}[3]{\mathchoice{\EXPhelp{#1}{#2}{#3}{\displaystyle}{<}{>}}{\EXPhelp{#1}{#2}{#3}{\textstyle}{<}{>}}{\EXPhelp{#1}{#2}{#3}{\scriptstyle}{<}{>}}{\EXPhelp{#1}{#2}{#3}{\scriptscriptstyle}{<}{>}}}
\newcommand{\EXP}[2]{\mathchoice{\EXPhelp{#1}{#2}{}{\displaystyle}{<}{>}}{\EXPhelp{#1}{#2}{}{\textstyle}{<}{>}}{\EXPhelp{#1}{#2}{}{\scriptstyle}{<}{>}}{\EXPhelp{#1}{#2}{}{\scriptscriptstyle}{<}{>}}}
\newcommand{\BRAhelp}[6]{\sbox{\expbox}{\ensuremath{#4#1}}\settowidth{\explength}{\rotatebox{90}{\ensuremath{#4\left#5\usebox{\expbox}\right#6}}}\ensuremath{#4\left#5\usebox{\expbox}\right#6_{{#4\hspace*{0.4ex}}\hspace*{-0.2\explength}#2}^{{#4\hspace*{0.4ex}}\hspace*{-0.2\explength}#3}}}
\newcommand{\BRA}[2]{\mathchoice{\BRAhelp{#1}{}{#2}{\displaystyle}{(}{)}}{\BRAhelp{#1}{}{#2}{\textstyle}{(}{)}}{\BRAhelp{#1}{}{#2}{\scriptstyle}{(}{)}}{\BRAhelp{#1}{}{#2}{\scriptscriptstyle}{(}{)}}}
\newbox{\atbox}
\newlength{\atlengtha}
\newlength{\atlengthb}
\newcommand{\AThelp}[4]{\sbox{\atbox}{\ensuremath{#3#1}}\settoheight{\atlengtha}{\ensuremath{#3\usebox{\atbox}}}\settodepth{\atlengthb}{\ensuremath{#3\usebox{\atbox}}}#3\addtolength{\atlengtha}{0.1ex}
#3\addtolength{\atlengthb}{0.75ex}\addtolength{\atlengtha}{\atlengthb}#1\rule[-\atlengthb]{0.1ex}{\atlengtha}_{\raisebox{0.12ex}{\ensuremath{\,#4#2}}}}\newcommand{\AT}[2]{\mathchoice{\AThelp{#1}{#2}{\displaystyle}{\scriptstyle}}{\AThelp{#1}{#2}{\textstyle}{\scriptstyle}}{\AThelp{#1}{#2}{\scriptstyle}{\scriptscriptstyle}}{\AThelp{#1}{#2}{\scriptscriptstyle}{\scriptscriptstyle}}}
\newlength{\picheight}\newcommand{\AxesPicture}[4]{\settowidth{\picheight}{\rotatebox{90}{$\tx#4$}}{\arraycolsep0.5ex
\renewcommand{\arraystretch}{1.3}
$\begin{array}{cc}
\mbox{\parbox{\picheight}{\rotatebox{90}{$~~\tx#4$}}}
 & \mbox{\epsfxsize=#2\textwidth\parbox{#2\textwidth}{\epsfbox{#1}}} \\
~ & \mbox{$~~\tx#3$}
\end{array}$}}
\newbox{\picbox}
\newlength{\picwidth}
\newcommand{\AxesPictureRight}[4]{\sbox{\picbox}{\epsfxsize=#2\textwidth\parbox{#2\textwidth}{\epsfbox{#1}}}
\settowidth{\picheight}{\rotatebox{90}{$\tx#4$}}\settowidth{\picwidth}{\rotatebox{90}{\usebox{\picbox}}}
\centerline{\arraycolsep0.5ex
\renewcommand{\arraystretch}{1.3}
$\begin{array}{cr}
\mbox{\parbox{\picheight}{\rotatebox{90}{\parbox{\picwidth}{\hfill{$\tx#4$}}}}}
 & \mbox{\usebox{\picbox}} \\
~ & \mbox{$\tx#3$}
\end{array}$}}
\newcommand{\AxesPictureSpecial}[6]{\centerline{$\raisebox{#5\textwidth}{\rotatebox{90}{$\tx#4$}}~
{{\epsfxsize=#2\textwidth\parbox{#2\textwidth}{\epsfbox{#1}}}}$$\hspace*{-2ex}\raisebox{#6\textwidth}{$\tx#3$}$}}
\newcommand{\NOhere}{\parbox{1cm}{\epsfxsize=1cm\epsfbox{./eps/isno-ani.eps}}}
\newcommand{\NO}{\marginpar{$\!\!\parbox{1.8cm}{\epsfxsize=1.8cm\epsfbox{./eps/isno-ani.eps}}$}}

\newcommand{\beq}{\begin{equation}}
\newcommand{\eeq}{\end{equation}}
\newcommand{\bea}{\begin{eqnarray}}
\newcommand{\eea}{\end{eqnarray}}
\newcommand{\bal}{\begin{align}}
\newcommand{\eal}{\end{align}}
\def\beginincorrect{

\noindent{\unitlength1mm
\begin{picture}(100,5)
\put(0,0){\line(0,1){5}}
\put(0,5){\line(1,0){179}}
\put(179,0){\line(0,1){5}}
\put(78,2){\mbox{\normalsize*****incorrect*****}}
\end{picture}}\newline
\scriptsize}
\def\endincorrect{

\noindent{\unitlength1mm
\begin{picture}(100,5)
\put(0,0){\line(0,1){5}}
\put(0,0){\line(1,0){179}}
\put(179,0){\line(0,1){5}}
\put(74,1){\mbox{\normalsize*****end incorrect*****}}
\end{picture}}

\normalsize}

\newcommand{\bilderscale}{0.35}
\newcommand{\textbilderscale}{0.25}

\newcommand{\fig}[2]{\includegraphics[width=#1]{#2}}
\newcommand{\pfig}[2]{\parbox{#1}{\includegraphics[width=#1]{#2}}}
\newcommand{\Fig}[1]{\includegraphics[width=\columnwidth]{#1}}
\newcommand{\rfig}[2]{\includegraphics[width=#1\columnwidth]{#2}}
\newlength{\bilderlength}
\newcommand{\usebilderscale}{\bilderscale}
\newcommand{\bilderskip}{\hspace*{0.8ex}}
\newcommand{\textdiagram}[1]{\renewcommand{\usebilderscale}{\textbilderscale}\diagram{#1}\renewcommand{\usebilderscale}{\bilderscale}}
\newcommand{\diagram}[1]{\settowidth{\bilderlength}{\bilderskip\includegraphics[scale=\usebilderscale]{#1}\bilderskip}\parbox{\bilderlength}{\bilderskip\includegraphics[scale=\usebilderscale]{#1}\bilderskip}}
\newcommand{\Diagram}[1]{\settowidth{\bilderlength}{\includegraphics[scale=\usebilderscale]{#1}}\parbox{\bilderlength}{\includegraphics[scale=\usebilderscale]{#1}}}

\newcommand{\LWC}{\cite{ChauveLeDoussalWiese2000a,LeDoussalWieseChauve2003}}

\newcommand{\ovl}[1]{\overline{#1}}
\newcommand{\mtin}[1]{\mbox{\tiny {#1}}}
\newcommand{\ca}[1]{{\cal #1}}
\newcommand{\sfrac}[2]{{\textstyle\frac{#1}{#2}}}

\title{\vspace{-1.5cm}\sffamily\Large\bfseries
Field Theory of Disordered Elastic Interfaces at 3-Loop Order: Critical Exponents and Scaling Functions}

\author{{\sffamily\bfseries Christoph Husemann\(^1\) and Kay J\"org
Wiese$^{2}$}\\
\small \(^{1}\)Carl Zeiss AG, Carl Zeiss Promenade 10, D-07745 Jena, Germany
\\ \small $^{2}$CNRS-Laboratoire de Physique Théorique de l'Ecole Normale Sup\'erieure, PSL Research University,\\ \small Sorbonne Universit\'es, UPMC, 24 rue Lhomond, 75005 Paris, France.}

\date{}\maketitle
\begin{abstract}
For disordered elastic manifolds in the ground state (equilibrium) we obtain the critical exponents for the roughness and the correction-to-scaling  up to 3-loop order, i.e.\ third order in $\epsilon=4-d$,  where $d$ is the internal dimension $d$.
 We also give the full  2-point function up to order $\epsilon^{2}$, i.e.\ at 2-loop order.
\end{abstract}

\section{Introduction}\label{intro}

For disordered system the application of the functional renormalization group (FRG) is non-trivial because of the cuspy form of the disorder correlator \cite{NarayanDSFisher1993a,NarayanDSFisher1992b,NarayanDSFisher1992a,LeschhornNattermannStepanowTang1997,NattermannStepanowTangLeschhorn1992,LeDoussalWieseChauve2003,LeDoussalWieseChauve2002,ChauveLeDoussalWiese2000a,MiddletonLeDoussalWiese2006}. In \cite{WieseHusemannLeDoussal2018} we obtained for a 1-component field ($N=1$) the $\beta$-function to 3-loop order, employing the  exact renormalization group and  several other techniques.  Here we analyze the fixed point: We calculate to 3-loop order the roughness exponent \(\zeta\) for random-bond disorder, the universal amplitude for periodic disorder, as well as the RG fixed-point functions and universal correction-to-scaling exponents. We also give the complete functional form of the universal 2-point function up to 2-loop order.

Our results are relevant for a remarkably broad set of problems, from
subsequences of random permutations in mathematics
\cite{Johansson2000}, random
matrices \cite{PraehoferSpohn2000a,PraehoferSpohn2000} to growth
models
\cite{KPZ,FreyTaeuber1994,Laessig1995,FreyTaeuberHwa1996,Wiese1997c,Wiese1998a,MarinariPagnaniParisi2000,PraehoferSpohn1997,Krug1997}
and Burgers turbulence in physics
\cite{Mezard1997,MedinaHwaKardarZhang1989}, as well as directed
polymers \cite{KPZ,HwaFisher1994b} and optimization problems such as
sequence alignment in biology
\cite{BundschuhHwa2000,BundschuhHwa1999,HwaLaessig1998}.  Furthermore,
they are very useful   for numerous experimental systems, each
with its specific features in a variety of situations.  Interfaces in
magnets
\cite{NattermannBookYoung,LemerleFerreChappertMathetGiamarchiLeDoussal1998}
experience either short-range disorder (random bond RB), or long range
(random field RF). Charge density waves (CDW) \cite{Gruner1988} or the
Bragg glass in superconductors
\cite{BlatterFeigelmanGeshkenbeinLarkinVinokur1994,GiamarchiLeDoussalBookYoung,GiamarchiLeDoussal1995,GiamarchiLeDoussal1994,NattermannScheidl2000}
are periodic objects pinned by disorder. The contact line of a
 meniscus on a rough substrate is governed by long-range
elasticity
\cite{LeDoussalWieseMoulinetRolley2009,PrevostRolleyGuthmann2002,PrevostThese,ErtasKardar1994b,LeDoussalWiese2009a}.  All
these systems can be parameterized by a $N$-component height or
displacement field $u(x)$, where $x$ denotes the $d$-dimensional
internal coordinate of the elastic object. An interface in the 3D random-field Ising model
has $d=2$, $N=1$, a vortex lattice $d=3$, $N=2$, a contact-line $d=1$
and $N=1$. The so-called directed polymer ($d=1$) subject to a short-range correlated disorder potential has been much
studied \cite{KardarLH1994} as it maps onto the Kardar-Parisi-Zhang
growth model \cite{KPZ,Krug1997,Wiese1998a} for any $N$, and yields an important check for the roughness exponent, defined below, \(\zeta_{\mathrm{eq,RB}}(d=1,N=1)=2/3\). Another important field of applications are avalanches, in magnetic systems known as Barkhausen noise. For applications and the necessary theory see e.g.\ \cite{AragonKoltonDoussalWieseJagla2016,DurinBohnCorreaSommerDoussalWiese2016,LaursonIllaSantucciTallakstadyAlava2013,DurinZapperi2000,PerkovicDahmenSethna1995,LeDoussalWiese2012a,DobrinevskiLeDoussalWiese2013,DobrinevskiLeDoussalWiese2011b,LeDoussalMuellerWiese2011,LeDoussalWiese2008a,ZhuWiese2017}.

Finally, let us note that the fixed points analyzed here are for equilibrium, a.k.a.\ ``statics''. At depinning, both the effective disorder, and the critical exponents change. A notable exception is periodic disorder and its mapping to loop-erased random walks \cite{WieseFedorenko2018}, where the disorder force-force correlator $\Delta(u)$ changes   by a constant, while all other terms are unchanged, and can be gotten from a simpler scalar field theory, allowing to extend the analysis done here to higher-loop order \cite{WieseFedorenko2018}.

\section{Model and basic definitions}

The equilibrium problem is
defined by the partition function ${\cal Z} := \int {\cal D}[u]\,
\exp(-{\cal H}[u]/T)$ associated to the Hamiltonian (energy)
\begin{equation} \label{ham}
 {\cal H}[u]= \int \rmd^d x\, \frac{1}{2} \left[\nabla u(x)\right]^2 +\frac{m^2}{2} \left[u(x)-w\right]^2 +V\big(u(x),x\big)
\ .
\end{equation}
In order to simplify notations, we will often note \begin{equation}\label{2.2}
\int_x f(x):=\int \rmd^d x\, f(x)\ ,
\end{equation}
and in momentum space
\begin{equation}\label{2.3}
\int_q \tilde f(q):=\int  \frac{\rmd^d q}{(2\pi)^d}\tilde f(q)\ .
\end{equation}
The Hamiltonian (\ref{ham}) is the sum of the elastic energy \(\int_x \frac{1}{2} \left[\nabla u(x)\right]^2 \) plus the confining potential  \(\frac{m^2}2 \int_x \left[u(x)-w\right]^2 \) which tends to suppress
fluctuations away from the  ordered state $u(x)={w}$, and
a random potential \(V(u,x)\) which enhances them. \(w\) is, up to a  factor of $m^2$, an applied external force, which is useful to measure the renormalized disorder  \cite{LeDoussal2006b,MiddletonLeDoussalWiese2006,LeDoussalWiese2006a,RossoLeDoussalWiese2006a,LeDoussalWieseMoulinetRolley2009,LeDoussalWiese2008a,LeDoussalMuellerWiese2007}, or properly define avalanches \cite{LeDoussalWiese2006a,RossoLeDoussalWiese2006a,LeDoussalWiese2009a,LeDoussalWiese2011b,LeDoussalMiddletonWiese2008,LeDoussalWiese2008c,RossoLeDoussalWiese2009a}.
The resulting roughness
exponent $\zeta$
\begin{equation}\label{lf5}
\overline{\left<[u(x) - u(x')]^2\right>} \sim |x-x'|^{2 \zeta}
\end{equation}
is measured in experiments for systems at equilibrium ($\zeta_{\rm
eq}$) or driven by a force $f$ at zero temperature (depinning, $\zeta_{\rm dep}$). Here and below $\left<\dots \right>$
denote thermal averages and $\overline {(\dots) }$
disorder ones. In the zero-temperature limit, the partition function is dominated by  the ground state, and  we may  drop the explicit thermal averages. In some cases, long-range elasticity appears, e.g.\ for
a contact line by integrating out the bulk-degrees of freedom
\cite{ErtasKardar1994b}, corresponding to $q^2 \to |q|$ in the elastic
energy. The random potential can without
loss of generality \LWC\ be chosen Gaussian with second cumulant \begin{equation}\label{corrstat}
\overline{V (u, x) V(u',x')} =: R_0(u-u') \delta^d(x-x') \ .
\end{equation}
\(R_0(u)\)  takes various forms: Periodic systems
are described by a periodic function $R_0(u)$, random-bond disorder by a short-ranged function, and random-field disorder of
variance $\sigma$ by $R(u) \simeq - \sigma |u|$ at large $u$.
Although this paper is devoted to equilibrium statics, some comparison
with dynamics will be made and it is thus useful to indicate the
corresponding equation of motion. Adding a time index to the field, \(u(x) \to u(x,t) \), the latter reads \begin{equation}\label{eqn.motion}
\eta \partial_t u(x,t) =-\frac{\delta{ \cal H}[u]}{\delta u(x,t)}= \nabla_x^2  u(x,t)  +m^2[w-u(x,t)]+ F\left(  u(x,t) ,x\right) \ ,
\end{equation}
with friction $\eta$. The (bare) pinning force is $F(u,x) = - \partial_u V(u,x)$, with correlator \begin{equation}
\Delta_0(u) = - R_0''(u)\ .
\end{equation}
To average over disorder, we replicate the partition function \(n\)
times,  \(\overline{{\cal Z}^n} =:\rme^{-\cal S}\), which defines the effective action \(\cal S\),
\begin{equation}
{\cal S}[u]= \sum_{a=1}^n \frac{1}{2T}\int_x \left[\nabla u_a (x)\right]^2
+ \frac{m^2}{2T}u_a(x)^2 -\frac1{2T^2}\int_x\sum_{a,b=1}^n R_0\big(u_{a}(x)-u_{b}(x)\big)\ .
\end{equation}
 We used the notations introduced in Eqs.~(\ref{2.2}) and (\ref{2.3}).
In presence of external sources \(j_a \), the \(n\)-times replicated action becomes
\begin{equation}\label{lf6}
{\cal Z}[j] := \int \prod_{a=1}^n {\cal D}[u_a]\, \exp\left(- {\cal S}[u] +
\int_x \sum_a j_a(x) u_a(x)\right) \ ,
\end{equation}
from which all static observables can be obtained.
$a$ runs from 1 to $n$, and the limit of zero  replicas $n\to 0$
is implicit everywhere.

\section{3-loop $\beta$-function}

In \cite{WieseHusemannLeDoussal2018} we derived the functional renormalization group equation for the renormalized, dimensionless disorder correlator \(\tilde R(u)\). For convenience we restate the $\beta$-function here
\begin{eqnarray}
-m\partial_{m} \tilde R(u) &=& (\epsilon-4\zeta) \tilde R(u) + \zeta u \tilde R'(u)+  \textstyle \frac{1}{2}
{\tilde R''(u)}^{2}-\tilde R''(u)\tilde R''(0)\nn \\
 && +\left({\textstyle \frac{1}{2}} + \epsilon\, {\cal C}_{1} \right)
\Big[\tilde R''(u) {\tilde R'''(u)}^{2}-\tilde R''(0) {\tilde R'''({u})}^{2} - \tilde R''(u) {
\tilde R'''({0^+})}^{2}  \Big] \nn \\
&& + {\cal C}_{2} \Big[ {\tilde R'''(u)}^4-2 {\tilde R'''(u)}^{2} {\tilde R'''(0^+)}^{2}
\Big]+ {\cal C}_{3}\,\big[\tilde R''(u)-\tilde R''(0)\big]^2 {\tilde R''''(u)}^2 \rule{0mm}{3ex} \nn \\
&&+ {\cal C}_{4}\, \Big[  \tilde R''(u) {\tilde R'''(u)}^{2}{\tilde R''''(u)}-\tilde R''(0)
{\tilde R'''(u)}^{2}{\tilde R''''(u)}
-\tilde R''(u) { \tilde R'''(0^+)}^{2} {\tilde R''''(0)}  \Big] \ .\qquad \ \ \
\label{betafinal}
\end{eqnarray}
The coefficients are \begin{eqnarray}
{\cal C}_{1} & =&\frac{1}{4} +\frac{\pi ^{2}}{9}-  \frac{\psi'
(\frac{1}{3})}{6}  = -0.3359768096723647... \label{C1}\\ \label{C2}
{\cal C}_{2}
&=& \frac{3}{4}\zeta (3)+\frac{\pi ^{2}}{18}-\frac{\psi'
(\frac{1}{3})}{12} = 0.6085542725335131...\\ \label{C3}
{\cal C}_{3}&=&  \frac{\psi'
(\frac{1}{3})}{6} -\frac{\pi ^{2}}{9}= 0.5859768096723648... \\
{\cal C}_{4} \label{C4}
&=& 2+\frac{\pi ^{2}}{9}-\frac{\psi' (\frac{1}{3})}{6}= 1.4140231903276352... \ .
\end{eqnarray}
The first line contains the rescaling and 1-loop terms, the second line the 2-loop terms, and the last two lines the three 3-loop terms. Note that \(\mathcal{C}_1=\frac14 -\mathcal{C}_3\), and \({\cal C}_4=2-{\cal C}_3 = \sqrt 2-0.000190372...  \)

\section{Summary of main results}
Here we summarize the main results. Their derivation is given in the following sections.

\subsection{Fixed points and critical exponents}
There are four  generic distinct  disorder classes, corresponding to random-bond, random-field, random-periodic, and generic long-ranged disorder. While we will discuss the details in section  \ref{FP-analysis}, we give a summary here.

\subsubsection{Random-bond disorder}

If the microscopic disorder potential is short-ranged, which corresponds to random-bond disorder in magnetic systems, then the roughness exponent can be calculated in an $\epsilon=4-d$ expansion:
\begin{align}
\zeta &= \epsilon \zeta_1 +\epsilon^2\zeta_2+\epsilon^3 \zeta_3 +\ca{O}(\epsilon^4)\\
\zeta_1 &= 0.2082980628(7)  \\
\zeta_2 &= 0.006857(8) \\
\zeta_3 &= -0.01075(2) \,.
\end{align}
This series expansion has a rather large third-order coefficient. As we will discuss in the conclusions, this is a little  surprising, since one might expect the expansion to converge, contrary to $\varphi^{4}$-theory which has a divergent, but Borel-summable series expansion.

 One can use a Pad\'e resummation  to improve the expansion. Asking that all Pad\'e coefficients are positive singles out  the  (2,1)-approximant. It is given by
\begin{align}
 \zeta_{(2,1)} \approx  \frac{0.208298 \epsilon + 0.333429 \epsilon^2}{1 +
1.56781 \epsilon} +\ca{O}(\epsilon^4) \ .
\end{align}
Adding a 4-loop term, and asking that in dimension one the exact result is reproduced, i.e.\ $\zeta(\epsilon=3)=2/3$, and choosing the Pad\'e with positive coefficients only, leads to \begin{equation}
\zeta \approx  \frac{0.0021794 \epsilon
   ^4+0.333429 \epsilon ^2+0.208298
   \epsilon }{1.56781 \epsilon +1}  +\ca{O}(\epsilon^4) \ .
\end{equation}
Details can be found in section \ref{s:RB-FP}.

\subsubsection{Random-field disorder}
The roughness exponent is given  \begin{equation}
\zeta_{\rm RF}=\frac{\epsilon}3 \ ,
\end{equation}
a result exact to all orders in $\epsilon$.
The amplitude of  the 2-point function can  be calculated analytically. It is given by
\begin{eqnarray}
&&  \overline{\left< \tilde u (q) \tilde u(q')\right>} = \tilde c(d)m^{-d-2 \zeta_{\rm RF}} F_d(q/m)\\ && F_d(0)=0\ ,\quad  F_d(z) \simeq B(d) z^{-d-2 \zeta_{\rm RF}} \mbox{~~for~~} z \to \infty \\
&&\tilde{c}(d) \approx \frac{\epsilon^{\frac 13} \sigma^{\frac  23}}{0.283721+
0.058367 \epsilon +
    0.064888 \epsilon^2}   + \ca{O}(\epsilon^{\frac{10}{3}}) \\
&& B(d)\approx \frac{1+ 0.226789 \epsilon}{1+ 0.560122 \epsilon} + \ca{O}(\epsilon^3)\ .
\end{eqnarray}
An analytical result is given in Eq.~(\ref{8.27})~ff.
We have again given the  Pad\'e approximants with only positive coefficients.
Translating to position space yields
\begin{equation}
\frac12 \overline{\left<[u(x)-u(0)]^2 \right>} = \frac{
   -\Gamma (-\tfrac \epsilon 3)\tilde c(d)B(d)}{(4\pi)^{\frac d2}\Gamma
   \left(\frac{d+8}{6} \right)} \left(\frac x 2\right)^{\!\!\tfrac{2 \epsilon}3}\ .
\end{equation}
The renormalization-group fixed point function \(R(u)\) for the disorder
can in this case be calculated analytically to third order in \(\epsilon\). The result, together with details on the   calculations is given in section \ref{s:RF-FP}.

\subsubsection{Periodic disorder}
For periodic disorder, the { 2}-point function is always a logarithm in position space, with universal amplitude, corresponding to \begin{equation}
\zeta_{\mathrm{RP}}=0\ .
\end{equation} The scaling functions are defined as for RF disorder, and read
\begin{eqnarray}
&&  \overline{\left< \tilde u (q) \tilde u(q')\right>} = \tilde c(d)m^{-d} F_d(q/m)\\ && F_d(0)=0\ ,\quad  F_d(z) \simeq B(d) z^{-d}
\mbox{~for~} z \to \infty \\
&&\tilde{c}(d) \approx \frac{2.19325 \epsilon}{1 + 0.310238 \epsilon + 1.33465
\epsilon^2} +\ca{O}(\epsilon^4)  \\
&& B(d) \approx \frac{1+ 0.134567 \epsilon  }{1+ 1.13457 \epsilon } + \ca{O}(\epsilon^3)\ .
\end{eqnarray}
An analytical result is given in Eq.~(\ref{9.21})~ff.
The  Pad\'e approximants are again given with only positive coefficients. Translating to position space yields, with a microscopic cutoff $a$
\begin{equation}
\frac12 \overline{ \left<[u(x)-u(0)]^2 \right> }= \frac{2
   \tilde c(d)B(d)}{(4\pi)^{\frac d2}\Gamma
    (\frac{d}{2}  )}  \ln |x/a|\ .
\end{equation}
Details are presented in sections \ref{s:scaling-function-for-aribtrary-zeta} and \ref{RP-FP}.

\subsection{Correction-to-scaling exponent}
The correction-to-scaling exponent $\omega$ quantifies how an observable $\ca O$, or a critical exponent, approaches its value at the IR fixed point at length scale $\ell$ or at mass $m$
\begin{equation}
\ca O -\ca O_{\rm fix-point} \sim \ell^{-\omega}  \sim m^{\omega}\ .
\end{equation}
For the fixed points studied above, the correction-to-scaling exponents are as follows.

\noindent
Random-Periodic fixed point:
\begin{equation}
\omega_{\rm RP} = -\epsilon +\frac{2 \epsilon ^2}{3}-\left(\frac{4 \zeta
   (3)}{3}+\frac{5}{9}\right) \epsilon ^3     +{\cal O}(\epsilon^{4})= -\epsilon\, \frac{1 + \left[2 \zeta (3)+\frac{1}{6}\right] \epsilon }{1+\left[ 2 \zeta (3)+\frac{5}{6}\right]
   \epsilon } +{\cal O}(\epsilon^{4})\ .
\end{equation}
Random-Bond fixed point:
\begin{equation}
\omega_{\rm RB}  \approx -\epsilon +0.4108  \epsilon^{2} +{\cal O}(\epsilon^{3})= -\frac{\epsilon }{1+0.4108  \epsilon }+{\cal O}(\epsilon^{3})\ .
\end{equation}
Random-Field fixed point:
\begin{equation}
\omega_{\rm RF} \approx -\epsilon +  0.1346 \epsilon ^{2} +{\cal O}(\epsilon^{3})= - \frac{\epsilon}{1+ 0.1346 \epsilon} +{\cal O}(\epsilon^{3})\ .
\end{equation}
Note that for the RP fixed point, we have given the solution up to 3-loop order. For the other fixed points, we have not attempted to solve the RG equations at this order, as this problem   can only  be tackled via shooting, which is already difficult at second order. Also, as the 2-loop result seems to be quite reliable, whereas   corrections for $\zeta$  are    large at 3-loop order,   we expect the same to be true for $\omega$, which  justifies to stop the expansion at second order.

Finally, we can perform the same analysis at depinning, with results as follows:

\noindent
Random-Field fixed point at depinning:
\begin{equation}
\omega_{\rm RF}^{\rm depinning}  \approx - \epsilon  - 0.0186\epsilon^{2} + {\cal O}(\epsilon^{3})
= -\frac{\epsilon }{1-  0.0186\epsilon} + {\cal O}(\epsilon^{3})\ .
\end{equation}
Random-Periodic fixed point at depinning: \begin{equation}
\omega_{\rm RP}^{\rm depinning} = - \epsilon  + \frac{2 \epsilon ^2}{3} +{\cal O}(\epsilon^{3})\ .
\end{equation}
Strangely, while the RP fixed point at depinning is different, the correction-to-scaling exponent $\omega$ does not change, at least to second order.

\subsection{2-point correlation function}

The 2-point correlation function can be written as
\begin{align}
{\overline{ \langle   u(q)   u(-q) \rangle} } = m^{-d-2\zeta} \tilde{c}(d) F_d(\sfrac{|q|}{m})\ ,
\end{align}
with a universal amplitude $\tilde{c}(d)$ and a scaling function $F_d$ with $F_d(0)=1$.
We have obtained the scaling function \(F_d(z)\) in an \(\epsilon \) expansion to second order, see Eqs.~\eq{Fd(z)-final} and \eq{8.27}.

\subsection{Other results}
A fixed-point function can also be constructed for generic long-ranged disorder, growing (or decaying) at large distances as \(R(u)\simeq u^\alpha\), with \(\alpha=1\) being random-field disorder discussed above. The idea is the same, in all cases the tail for large $u$ does not get corrected.

\section{Fix-point analysis}\label{FP-analysis}
Irrespective of the precise form of the initial disorder distribution function $R_0$ in the bare action, we identify different fix-point classes of the RG equation. Although our description may not be complete, the analysis of fix-point solutions gives insight into possible physical realizations of our simple model [Eq.~(\ref{ham})].
We study a universality class where $R_0$ is periodic and in the non-periodic case we distinguish whether $R_0$ is short range (random bond disorder) or long range (random field disorder). This chapter follows closely Ref. \cite{LeDoussalWieseChauve2003} but generalizes the results to 3-loop order.

In terms of the  rescaled disorder distribution function
\begin{align}\label{R-rescale}
 \tilde{R}(u) = (\epsilon \tilde{I}_1) m^{-\epsilon+4\zeta} R(um^{-\zeta})
\end{align}
the $\beta$-function up to 3-loop order was given in \Eq{betafinal}.

\subsection{Random-bond disorder}\label{s:RB-FP}
In order to describe short-range disorder caused by random bonds we look for a fix-point solution that decays exponentially fast for large fields $u$. To this end we numerically solve the fix-point equation
\begin{align}
-m \frac{\partial}{\partial m} \tilde{R}(u) =0
\end{align}
order by order in $\epsilon$. We make the Ansatz
\begin{align}
 \tilde{R}(u)& = \epsilon r_1(u) +\epsilon^2 r_2(u)+\epsilon^3 r_3(u) + \ca{O}(\epsilon^4)
\end{align}
and assume that higher orders in $\epsilon$ do not contribute to field derivatives of lower orders.
Also the roughness exponent is expanded in $\epsilon$
\begin{align}
\zeta &= \epsilon \zeta_1 +\epsilon^2\zeta_2+\epsilon^3 \zeta_3 +\ca{O}(\epsilon^4)\, .
\end{align}
If $\tilde{R}(u)$ is a fix-point solution of Eq.~(\ref{betafinal}) then $\xi^4\tilde{R}(u/\xi)$ is a fix-point solution as well for any $\xi$.
Thus, without loss of generality, it is possible to normalize $\tilde{R}(0)=\epsilon$, that is, we set $r_1(0)=1$ and $r_2(0)=r_3(0)=0$.

Inserting the Ansatz into the fix-point equation we find to lowest, that is, second order in $\epsilon$
\begin{align}
 0=(1-4\zeta_1)r_1(u) +\zeta_1ur_1'(u) +\frac 12 r_1''(u)^2-r_1''(u)r_1''(0)\,.
\end{align}
Together with $r_1(0)=1$ this differential equation has a solution for any $\zeta_1$. But for only one specific value of $\zeta_1$ the solution does not change sign and decays exponentially fast for large $u$. Since $R(u)=R(-u)$ we only consider positive values of $u$.

Numerically, we adopt the following iterative procedure: First, we guess a value for $\zeta_1$ and compute the corresponding $r_1(u)$. Then we evaluate $r_1$ at a large value $u_{\max}$. This is repeated until $r_1(u_{\max})= 0$. The guessing of $\zeta_1$ is improved by calculating $r_1(u_{\max})$ for many values of $\zeta_1$ and interpolating to zero. In order to circumvent numerical problems at small $u$ we approximate $r_1(u)$ by its Taylor expansion up to a finite order for $|u|$ smaller than a gluing point $u_{\mtin{glue}}$. We find
\begin{align}
 \zeta_1 &= 0.2082980628(7)
\end{align}
for $u_{\max}=25$. Below $u_{\mtin{glue}}=0.01$  a Taylor expansion of order 30 was used. The result does not depend on the order if high enough, also a reasonable variation of the gluing point $u_{\mtin{glue}}<2$ is within error tolerances (that is, does not change the digits shown here). Of course, the result does depend on $u_{\max}$, but choosing $u_{max}>25$ gives results within error tolerances (checked up to $u_{\max}=50$).

\begin{figure}
\begin{center}
\begin{minipage}{0.49\textwidth}
\includegraphics[width=0.9945\textwidth,angle=0]{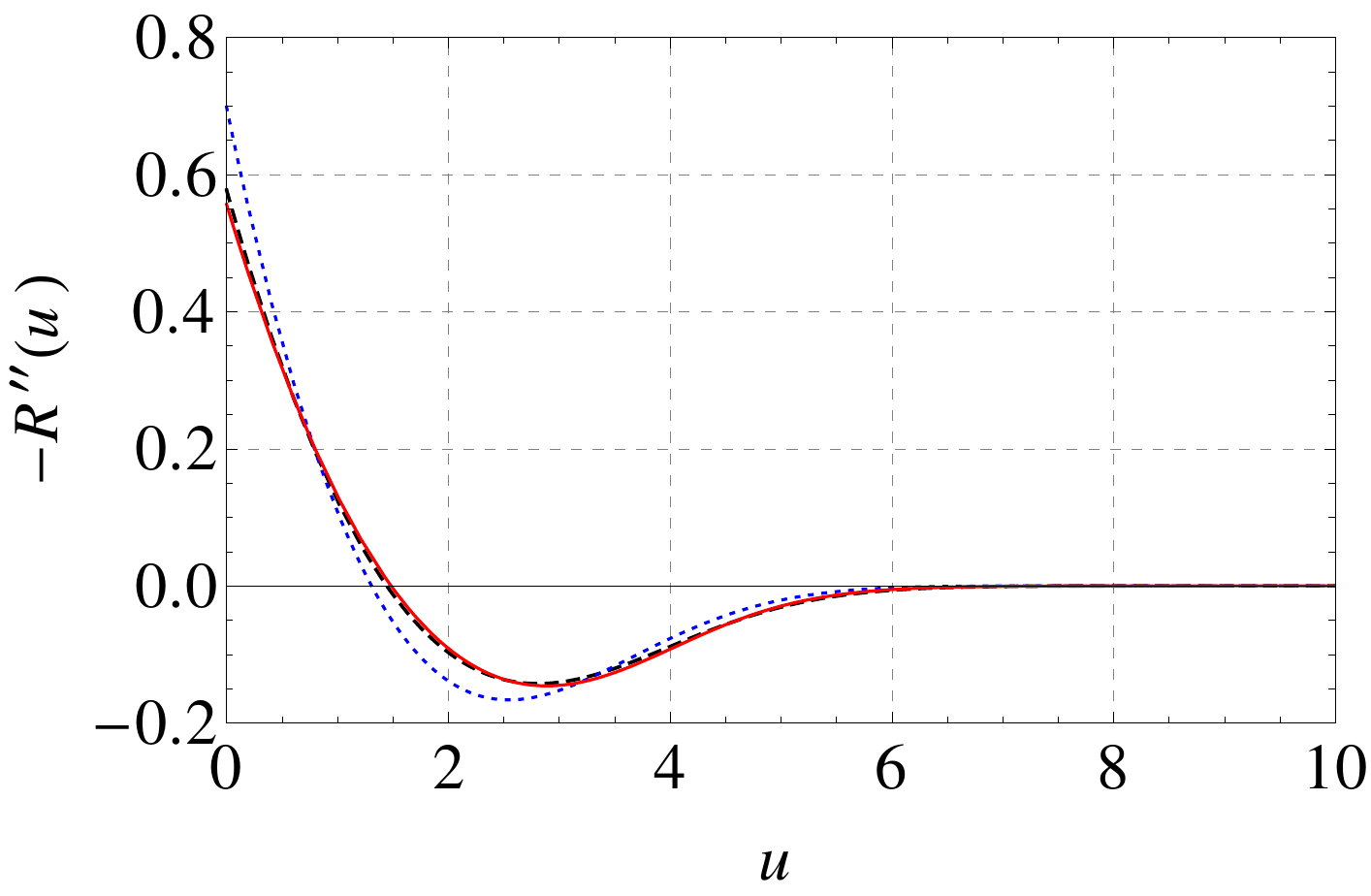}
\end{minipage}
\begin{minipage}{0.49\textwidth}
\includegraphics[width=0.9945\textwidth,angle=0]{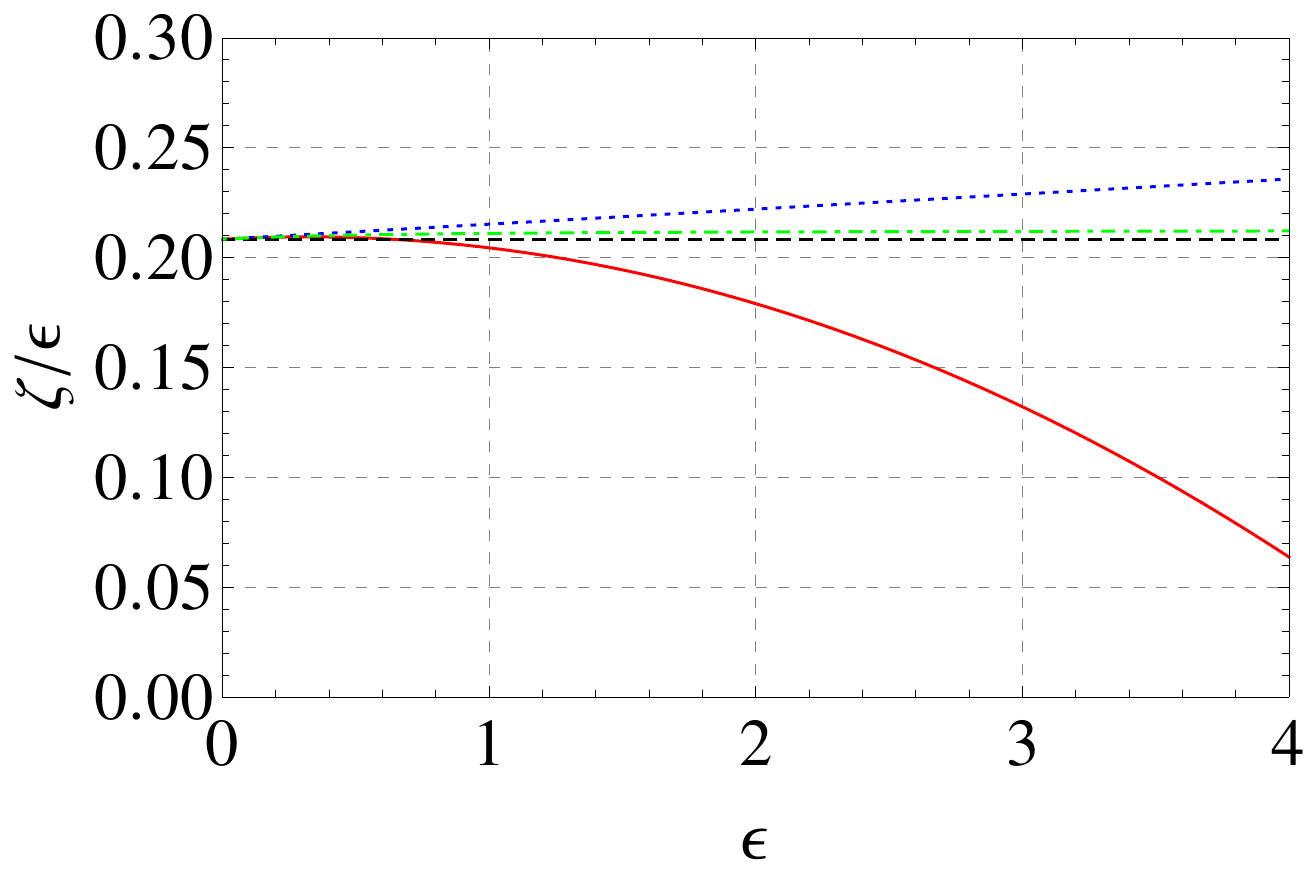}
\end{minipage}
\caption{Comparison of results for random bond disorder in 1-loop (black, dashed), 2-loop (blue, dotted), and 3-loop (red, solid) order. Left: Fix point disorder correlator for $d=3$. Right: Dimensional dependence of the roughness exponent normalized with $\epsilon$. The (2,1)-Pad\'e approximant is plotted in a green dash-dotted line.} \label{fig:RBfixpointfunction}
\end{center}
\end{figure}

For the higher-loop contributions we also need derivatives of $r_1$. Instead of solving the corresponding differential equations we simply take numerical derivatives. This is possible since $r_1$ is a smooth function away from zero.
Using the obtained $\zeta_1$, $r_1(u)$, and its derivatives, we can solve for the 2-loop contribution
\begin{align}
0&=  r_{2} (u) - 4 \zeta_{2}r_{1} (u)  -4 \zeta_{1} r_{2} (u) + u
\zeta_{2}r_{1}' (u) + u \zeta_{1} r_{2}' (u) + r_{1}'' (u)r_{2}'' (u)
-  r_{1}'' (0)r_{2}'' (u)  \\ \nonumber & \quad - r_{1}'' (u)r_{2}'' (0)
 + \frac{1}{2} \left(r_{1}'' (u)-r_{1}'' (0) \right) r_{1}'''
(u)^{2} -\frac{1}{2} r_{1}'' (u)r_{1}''' (0^{+})^{2}
\end{align}
with $r_2(0)=0$. This equation is solved for $r_2(u)$ for different values of $\zeta_2$. With an analogous iterative procedure we adjust $\zeta_2$ such that $r_2(u)$ decays exponentially. The best value is
\begin{align}
 \zeta_2 &= 0.006857(8)
\end{align}
as found in \cite{LeDoussalWieseChauve2003}. Again, derivatives of $r_2(u)$ are computed numerically and put into the 3-loop contribution
\begin{align}
 0&= r_3(u)-4 \zeta_3 r_1(u)-4 \zeta_2 r_2(u)-4 \zeta_1 r_3(u)+u \big[\zeta_3 r_1'(u)+ \zeta_2 r_2'(u)+ \zeta_1 r_3'(u)\big] \\ & \nonumber \quad -r_2''(0) r_2''(u)+\sfrac{1}{2} r_2''(u)^2-r_1''(u) r_3''(0)-r_1''(0) r_3''(u)+r_1''(u) r_3''(u)-\ca{C}_1 r_1''(u) r_1'''(0)^2 \\ & \nonumber \quad - \sfrac{1}{2} r_2''(u) r_1'''(0)^2-\ca{C}_1 r_1''(0) r_1'''(u)^2 + \ca{C}_1 r_1''(u) r_1'''(u)^2- \sfrac{1}{2} r_2''(0) r_1'''(u)^2+ \sfrac{1}{2} r_2''(u) r_1'''(u)^2  \\ &\nonumber \quad - 2 \ca{C}_2 r_1'''(0)^2 r_1'''(u)^2+ \ca{C}_2 r_1'''(u)^4-r_1''(u)  r_1'''(0) r_2'''(0)- r_1''(0) r_1'''(u) r_2'''(u)+ r_1''(u) r_1'''(u) r_2'''(u) \\ &\nonumber \quad - \ca{C}_4 r_1''(u) r_1'''(0)^2 r_1''''(0)- \ca{C}_4 r_1''(0) r_1'''(u)^2 r_1''''(u) + \ca{C}_4 r_1''(u) r_1'''(u)^2 r_1''''(u) + \ca{C}_3 r_1''(0)^2 r_1''''(u)^2  \\ &\nonumber \quad- 2 \ca{C}_3 r_1''(0) r_1''(u) r_1''''(u)^2+ \ca{C}_3 r_1''(u)^2  r_1''''(u)^2
\end{align}
with normalization $r_3(0)=0$. With the iterative procedure described above an approximate exponential decay of $r_3(u)$ is found for
\begin{align}
 \zeta_3 &= -0.01075(2) \,.
\end{align}
The force correlator $-R''(u)$ of the fix-point is plotted on the left side in Fig.~\ref{fig:RBfixpointfunction} for $d=3$, that is, $\epsilon=1$ in a one-, two-, and 3-loop approximation. There are further renormalizations of the cusp, in particular, the 3-loop contribution seems to counteract the 2-loop contribution such that the 3-loop results is close to the 1-loop result.

The dimensional dependence of the roughness exponent is shown in the right graph of Fig.~\ref{fig:RBfixpointfunction}. The corrections in 3-loop order are substantial, for $\epsilon>1$ they are so large that the $\epsilon$-expansion is bound to fail. Correspondingly, while the 2-loop results seem to reproduce exact ($d=1$) \cite{KardarHuseHenleyFisher1985} and simulation results ($d=2,3$) \cite{Middleton1995}, the 3-loop results are worse throughout in this comparison, see Fig.~\ref{tab:RoughnessRB}. Surprisingly, the (2,1)-Pad\'e-approximant of the 3-loop $\epsilon$-expansion, which is given by
\begin{align}
 \zeta_{(2,1)}\approx \frac{0.208298 \epsilon + 0.333429 \epsilon^2}{1 + 1.56781 \epsilon},
\end{align}
is again very close to the 1-loop result but agrees even better with the reference data. Unfortunately, the third order of a series does not allow to make statements of its asymptotic behavior.

\begin{figure*}[bt]
\begin{center}
\begin{tabular}{||c|c|c|c|c|c||}
\hline
$\zeta _{\rm eq}$ & one loop & two loop & three loop & Pad\'e-(2,1) &
simulation and exact\\
\hline \hline $d=3$  & 0.208 &  0.215  & 0.204  &
0.211 &
$0.22\pm 0.01$ \cite{Middleton1995}  \\
\hline
$d=2$ &0.417 &0.444 & 0.358 & 0.423 & $0.41\pm 0.01$ \cite{Middleton1995} \\
\hline
$d=1$ & 0.625 & 0.687 & 0.396 & 0.636 & 2/3 \cite{KardarHuseHenleyFisher1985} \\
\hline
\end{tabular}
\vspace{1mm}
\caption{Roughness exponent for random bond disorder obtained by an $\epsilon$-expansion in comparison with exact results and numerical simulations. In the fourth column is an estimate value using a (2,1)-Pad\'e approximant of the 3-loop result.}
\label{tab:RoughnessRB}
\end{center}
\end{figure*}

\subsection{Random-field disorder}\label{s:RF-FP}
We consider a class of long-range fix-point solutions with $\tilde{R}(u) \sim -\sigma |u|$ for large $u$. Due to the linear behavior, second and higher derivatives of $\tilde{R}$ do not contribute in the limits $u\to \pm\infty$. Subsequently, all loop corrections to the tail vanish and the reparameterisation terms give $\zeta=\frac{\epsilon}{3}$ for the roughness exponent to all orders as a prerequisite for the existence of such a fix-point solution.

\begin{figure}
\begin{center}
\includegraphics[width=0.6\textwidth,angle=0]{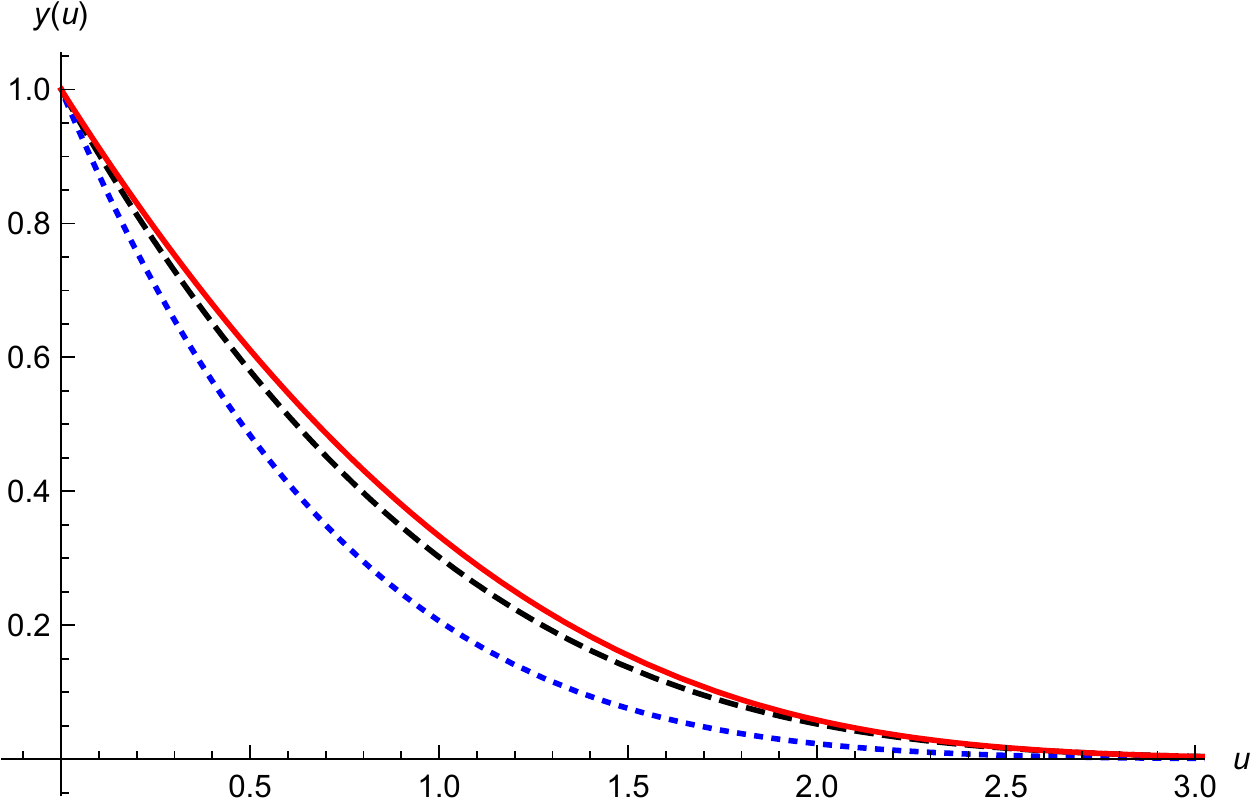}
\caption{Fix-point solution $y(u)= -\frac{3}{\epsilon} \tilde{R}''(u)$ for $\epsilon =1$ in the case of random-field disorder. Comparison of 1-loop (dashed black), 2-loop (blue dotted), and 3-loop (red line).} \label{fig:RFfixpoint}
\end{center}
\end{figure}

Following closely the 2-loop calculation \cite{LeDoussalWieseChauve2003}, we consider $y(u)=-\frac{3}{\epsilon} \tilde{R}''(u)$ and normalize $y(0)=1$. Rewriting the fix-point equation in terms of $y$ and integrating over the interval $[0^+,u]$ gives (without loss of generality we consider $u>0$)
\begin{align}\label{eq:RBfixpointeq}
 0&= B_1 + B_2 \epsilon +B_3 \epsilon^2 + \ca{O}(\epsilon^4)\ .
\end{align}
This is equivalent to taking one derivative of the $\beta$ function (\ref{betafinal}), and expressing
it in terms of $y$. The coefficients  $B_1$, $B_2$ and $B_3$ are  the 1-, 2-, and 3-loop contributions,  given by
\begin{align}\label{B1}
B_1 =& uy +(1-y)y' ,\\
B_2 =& \frac 16 \Big[ y'^2(y-1) \Big]' - \frac 16 y' y'(0)^2
\label{B2}
\end{align}
and
\begin{align}
B_3 =& \frac 19 y'(0)^2 \Big[ (C_4 y''(0) -3 C_1) y +2C_2 y'^2\Big]' \nn\\
& +\frac 19 \Big[ -C_2y'^4 { -}C_3(1-y)^2 y''^2 + (1-y)y'^2 (C_4 { y''} -3C_1) \Big]'\ .
\end{align}
These equations can be solved analytically, expressing $u$ as a function of $y$.
For the 1-loop equation, the solution reads
\begin{align} \frac{u^2}{2} = y -1- \ln y ,
\end{align}
which features the cusp.
Higher-loop contributions are obtained by making an ansatz; to 3-loop order we need \begin{align}\label{eq:RBansatz}
 \frac{u^{2}}{2} = y -1- \ln y - \frac{\epsilon}{3} F_2(y) - \frac{\epsilon^2}{6} F_3(y) + \ca{O}(\epsilon^3) \ , \qquad u>0\ .
\end{align}
The    inverse function of $u(y)$, $u>0$,  is $y(u)$. We make use of the known 2-loop solution \cite{LeDoussalWieseChauve2003}
\begin{align}
 F_2(y) =2y -1- \frac 12 \ln y  +\frac{y}{1-y} \ln y + \mbox{Li}_2(1-y)
\end{align}
with boundary conditions up to 2-loop order
\begin{align}
 y'(0) &= -1 -\frac 29 \epsilon  + \ca{O}(\epsilon^2)  \\ \nonumber
 y''(0) &= \frac 23 +\frac{19}{54} \epsilon + \ca{O}(\epsilon^2)  \\ \nonumber
 y'''(0) &= -\frac 16 -\frac{71}{360} \epsilon + \ca{O}(\epsilon^2) \ .
\end{align}
Differentiating the ansatz (\ref{eq:RBansatz}), with respect to $u$ gives an $\epsilon$-expansion for $y'(u)$
\begin{align}\label{eq:RBepsilonyprime}
y'(u) = - \frac{u y(u)}{1-y(u)} -\frac 13 \frac{y(u)y'(u)}{1-y(u)} \left[ \epsilon \frac{\mathrm{d}}{\mathrm{d}y} F_2(y)\Big|_{y=y(u)}   + \frac 12 \epsilon^2 \frac{\mathrm{d}}{\mathrm{d}y} F_3(y)\Big|_{y=y(u)}  \right] + \ca{O}(\epsilon^3)\ .
\end{align}
We now insert this expression into Eq.~(\ref{eq:RBfixpointeq}), replacing for the moment only $B_{1}$ by its explicit form (\ref{B1}). Then the fix-point condition reads
\begin{align}\label{eq:RBfixpointhelp}
 0= \epsilon \left[ B_2 - \frac 13 y(u)y'(u)F_2'(y(u)) \right]  +\epsilon^2 \left[ B_3-\frac 16 y(u) y'(u) F_3'(y(u))\right] + \ca{O}(\epsilon^3)\ .
\end{align}
The two terms, each enclosed by square brackets, are dealt with separately. We integrate the first term with respect to $u$ and then again insert Eq.~(\ref{eq:RBepsilonyprime}) to shift the occurrence of $y'(u)$ to a higher order in $\epsilon$. Since $F_2$ determines the 2-loop fixed point, the expression is of order $\epsilon$
\begin{align}
 \int_{0^+}^{\ovl{u}} \mathrm{d} u\left[ B_2 - \frac 13 y(u)y'(u)F_2'(y(u)) \right] =: \uu_1(y(\ovl{u}),\ovl{u}) \epsilon +\ca{O}(\epsilon^2).
\end{align}
The function $F_3$ can now be determined by considering the $\epsilon^2$-contribution to Eq.~(\ref{eq:RBfixpointhelp}),
\begin{align}
 B_3 - \frac 16 F_3'(y(u)) y'(u) y(u) + \frac{\mathrm{d}}{\mathrm{d} u} \uu_1(y(u),u) =0
\end{align}
and $F_3(1)=0$.
Dividing by $y(u)$ and integrating over $u$ we find
\begin{align}
 F_3(y(u)) = 6 \int^u \mathrm{d} u \frac{1}{y(u)} \left[ B_3 +  \frac{\mathrm{d}}{\mathrm{d} u} \uu_1(y(u),u)\right]\ .
\end{align}
(The lower bounds from the left and right-hand side cancel.) The integral on the right-hand-side is evaluated by first integrating
\begin{align}
\Psi(u)=\int^u \mathrm{d} u \left[ B_3 +  \frac{\mathrm{d}}{\mathrm{d} u} \uu_1(y(u),u)\right]
\end{align}
and then replacing $y'(u)$ and $y''(u)$ by Eq.~(\ref{eq:RBepsilonyprime}) and $u^2$ by Eq.~(\ref{eq:RBansatz}) to zeroth order in $\epsilon$. The remaining integral
\begin{align}
 \int\mathrm{d} u \frac{1}{y(u)} \frac{\mathrm{d}}{\mathrm{d}u} \Psi(u) =\tilde{F}_3(y(u))
\end{align}
can  be evaluated with the help of Mathematica and is a function of $y(u)$ only. We find
\begin{align}
 F_3(y)= \tilde{F}_3(y) - \tilde{F}_3(1) = f_0  + f_1 \ln y + f_2 (\ln y)^2 +f_3 \ln y \ln( 1-y) ,
\end{align}
where
\begin{align}
 f_0 &= \frac{4 + \frac{4 \pi^2}{9}- \frac 23 \gamma_{\frac{1}{3}} - 2 \zeta(3)}{(1 - y)^2}  -\frac{48 - \pi^2 - 30 \zeta(3)}{9 (1 - y)}   -\frac{1}{18} \Big[6 - 3 \gamma_{\frac{1}{3}} + (57 + 8 \pi^2) \zeta(3)\Big]\nonumber \\  &
~~~+ (1 - y) \Big[-\frac 49 \pi^2 + \frac 23  + \frac 23 \gamma_{\frac{1}{3}}  + 4 \zeta(3)\Big]
 +\frac 23 \left( 4 \zeta(3)-  \frac{ 2 - y }{1 - y}\right)  \mbox{Li}_2(y) -  \frac 23 \mbox{Li}_3(1-y)
\end{align}
with $\gamma_{\frac{1}{3}} =\psi'(\sfrac{1}{3})$. Furthermore,
\begin{align}
f_1 &= \frac{23}{18} - \frac{38 \pi^2}{81} + \frac{19}{27} \gamma_{\frac{1}{3}}  + \frac{4}{27}\frac{
 -2 \pi^2 + 3 (9 + \gamma_{\frac{1}{3}} - 15 \zeta(3))}{
 1 - y} -
 \frac{2}{27} \frac{180 + 4 \pi^2 - 6 \gamma_{\frac{1}{3}} - 117 \zeta(3)}{(1 - y)^2}\\ \nonumber &
    ~~~+ \frac{
 8 + \frac{8 \pi^2}{9} - \frac 43 \gamma_{\frac{1}{3}} - 4 \zeta(3)}{(1 - y)^3} +
 2 \zeta(3)  \\
f_2 &= \frac{4 + \frac{4 \pi^2}{9} - \frac 23  \gamma_{\frac{1}{3}} - 2 \zeta(3)}{(1 - y)^4}
 -  \frac{8 + \frac{8\pi^2}{27} - \frac 49 \gamma_{\frac{1}{3}} - \frac{16}{3} \zeta(3)}{(1 - y)^3} +
\frac{4 - \frac{4 \pi^2}{27} + \frac 29 \gamma_{\frac{1}{3}} - 4 \zeta(3)}{(1 - y)^2}  +\frac 23 \zeta(3) \\
f_3 &=  -\frac 23 - \frac{\frac 23}{1 - y} + \frac 83 \zeta(3)\ .
\end{align}
The functions $F_2$ and $F_3$ correct the cusp without destroying it, since both have a finite Taylor expansion around $y=1$,
\begin{align}
 F_2(y) =& \,\frac 23  (1 - y)^2 + \frac{13}{36} (1 - y)^3 + \frac{19}{80} (1 - y)^4 +\frac{13}{75} (1-y)^5
 +\frac{17}{126}
   (1-y)^6+\frac{43}{392} (1-y)^7\nn\\
   &\,+\frac{53}{576}
   (1-y)^8+\frac{32}{405} (1-y)^9+\frac{19}{275}
   (1-y)^{10}+\ca O(1-y)^{11} \\
 F_3(y) =&\, -2.08216 (1-  y)^2 - 0.949217 (1-  y)^3 -
 0.541283 (1-  y)^4-0.350724  (1-y)^5 \nn\\
 & \,- 0.247215 (1-y)^6- 0.185059
   (1-y)^7-0.144938 (1-y)^8-0.117575 (1-y)^9 \nn\\
   &\, -0.0980832
   (1-y)^{10} + \ca O (1-y)^{11}\ .
\end{align}
Both Taylor-expansions seem to be convergent in the whole range of $y$.
The 3-loop contribution has the opposite sign as the 2-loop contribution. For $\epsilon=1$ the 3-loop result corrects the 1-loop result in a different direction than the 2-loop result, see Fig.~\ref{fig:RFfixpoint}. The  3-loop contribution is larger than the 2-loop contribution, and the 3-loop result is closer to the 1-loop result.

\begin{figure}
\begin{center}
\includegraphics[width=0.5\textwidth,angle=0]{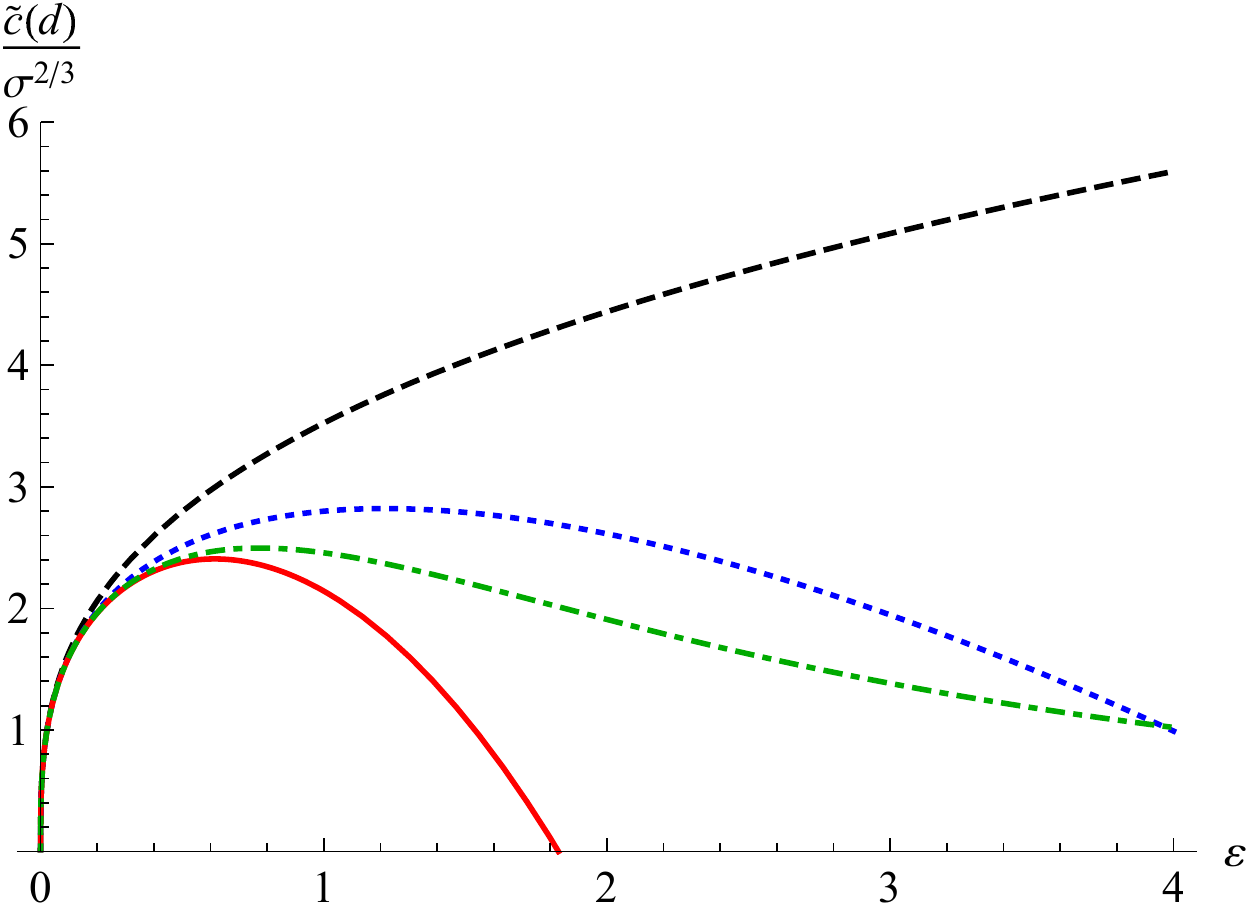}
\caption{Dimensional dependence of the universal amplitude for random-field disorder. Comparison of 1-loop (dashed black), 2-loop (blue dotted), and 3-loop (red line). The green dot-dashed line is the  (0,2)-Pad\'e approximant of the 3-loop solution.}
\label{fig:univAmplitudeRFfixpoint}
\end{center}
\end{figure}

If $\tilde \Delta(u):=-\tilde{R}''(u)=\frac{\epsilon}{3} y(u)$ is a fix-point solution, then
\begin{align}\label{eq:reparaminv}
 \tilde \Delta(u)= -\tilde{R}''(u)=\frac{\epsilon}{3}\xi^2 y(\sfrac{u}{\xi})
\end{align} is a fix-point solution for any $\xi$ as well. We choose $\xi$ to set the normalization of the fix-point function such that $\tilde{R}(u)\sim -\tilde{\sigma} |u|$  for large $u$, where $\tilde{\sigma}=(\epsilon \tilde{I}_1) \sigma$. This ensures $R(u)=\frac{1}{\epsilon \tilde{I}_1} m^{\epsilon-4\zeta} \tilde{R}(um^\zeta) \sim -\sigma |u|$, with $\zeta=\frac\epsilon3$ . The constant $\xi$ is determined by
\begin{align}
 \tilde{\sigma} \overset{!}{=} -\int_0^{\infty} \mathrm{d} u \; \tilde{R}''(u) = \frac{\epsilon}{3} \xi^3 \int_0^\infty \mathrm{d}u \; y(u) =\frac{\epsilon}{3} \xi^3 \int_0^1\mathrm{d}y \;  u(y)=\frac{\epsilon}{3} \xi^3 {\ca I_y}\ .
\end{align}
The (implicit) solution $u(y)$ is given by the ansatz~(\ref{eq:RBansatz}). Numerically, the integral is given by
\begin{align}
 {\ca I_y}&=\int_0^1\mathrm{d}y \;  u(y)  \approx 0.775304 -0.139455 \epsilon+ 0.17420\epsilon^2+ \ca{O}(\epsilon^3)\ .
\end{align}
With this fix-point solution, we  calculate the universal amplitude as
\begin{align}
 \tilde{c}(d) = m^{d+2\zeta} \langle u(0) u(0) \rangle = -\frac{1}{ \epsilon \tilde{I}_1} \tilde{R}''(0) = \frac{1}{(\epsilon \tilde{I}_1)^{\frac{1}{3}}} \left(\frac{\epsilon}{3}\right)^{\frac{1}{3}} \sigma^{\frac{2}{3}} {\ca I_y}^{-\frac 23}\ .
\end{align}
Using formulas (\ref{N})--(\ref{epsI1}),
we obtain
\begin{align}
 \tilde{c}(d) \approx \epsilon^{\frac 13} \sigma^{\frac 23} \Big[ 3.52459  - 0.72508 \epsilon -
 0.65692 \epsilon^2 + \ca{O}(\epsilon^3) \Big]\ .
\end{align}
For $\epsilon<0.5$ the 3-loop solution is relatively close to the 2-loop contribution. For larger $\epsilon$ it deviates substantially and even changes sign for $\epsilon \approx 1.83$, see Fig.\ref{fig:univAmplitudeRFfixpoint}.

The comparison with the exact result in $d=0$ dimensions \cite{LeDoussalMonthus2003} may be far fetched in an $\epsilon=4-d$ expansion. The 1- and 3-loop results are far off from the exact result, but the 2-loop result comes surprisingly close, see Fig.~\ref{tab:univAmplitudeRF}. More convincingly and even closer to the exact result is   the { (0,2)}-Pad\'e approximant of the 3-loop result, taking out the prefactor of $\epsilon^{1/3}$, which is the unique  approximant with only positive coefficients,
\begin{align}
\tilde{c}(d)_{ (0,2)} \approx \frac{\epsilon^{\frac 13} \sigma^{\frac  23}}{0.283721+ 0.058367\epsilon +
    0.064888 \epsilon^2}\ .
\end{align}\begin{figure*}[bt]
\begin{center}
\begin{tabular}{||c|c|c|c|c|c||}
\hline
$\tilde{c}(d) \sigma^{-\frac{2}{3}}$ & one loop & two loop & three loop & Pad\'e-(0,2) &
exact\\
\hline
\hline $d=3$  & 3.525 & 2.800   & 2.143 & 2.457 &   \\
\hline $d=2$  & 4.441 & 2.614   & -0.697 & 1.909 &   \\
\hline $d=1$  & 5.083 & 1.946   & -6.581 & 1.383 &  \\
\hline $d=0$  & 5.595 &  0.991  & -15.694  & 1.021 & $\approx 1.054 $ \cite{LeDoussalMonthus2003} \\
\hline
\end{tabular}
\vspace{1mm} \caption{Universal amplitude for random field disorder obtained by an $\epsilon$-expansion in comparision with the exact result. In the fourth column is the estimated value using a (0,2)-Pad\'e approximant of the 3-loop result.}
\label{tab:univAmplitudeRF}
\end{center}
\end{figure*}

\subsection{Periodic systems} \label{RP-FP}
\begin{figure}
\begin{center}
\includegraphics[width=0.5\textwidth,angle=0]{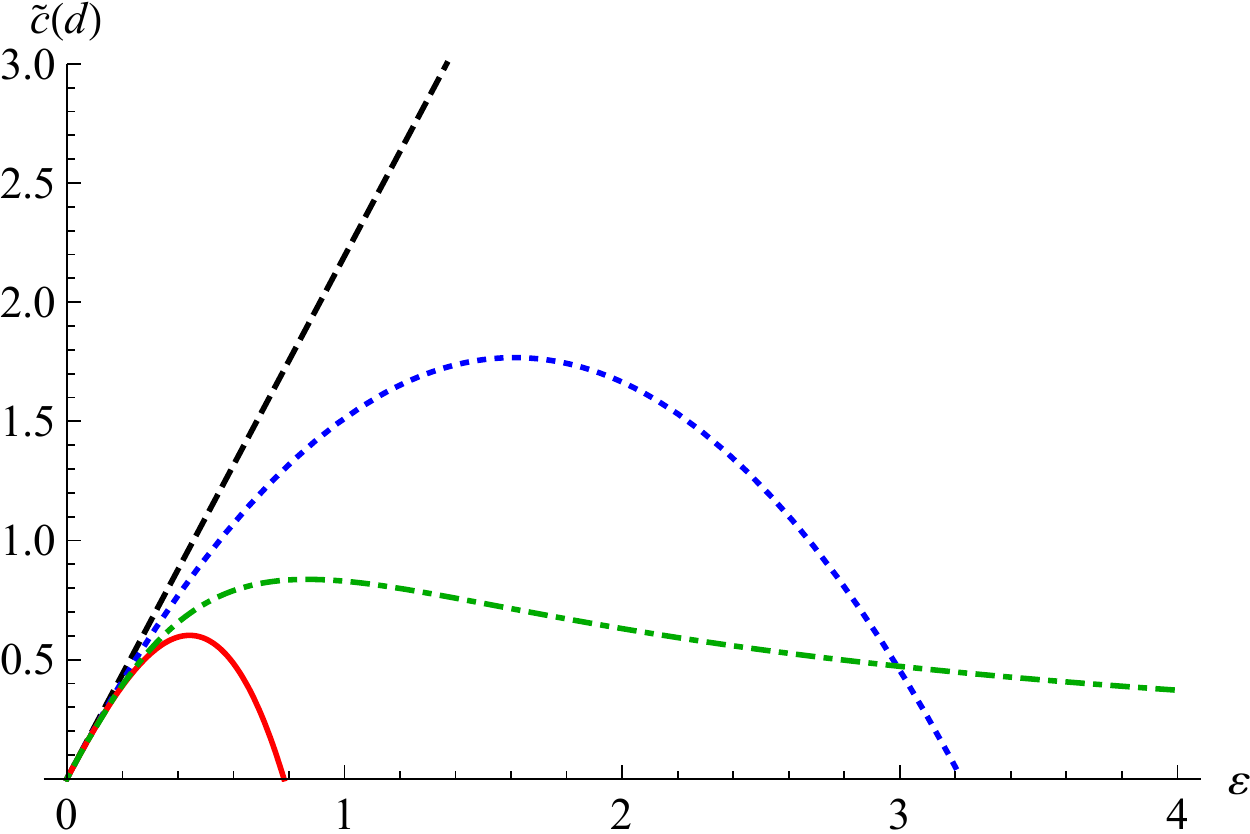}\includegraphics[width=0.5\textwidth,angle=0]{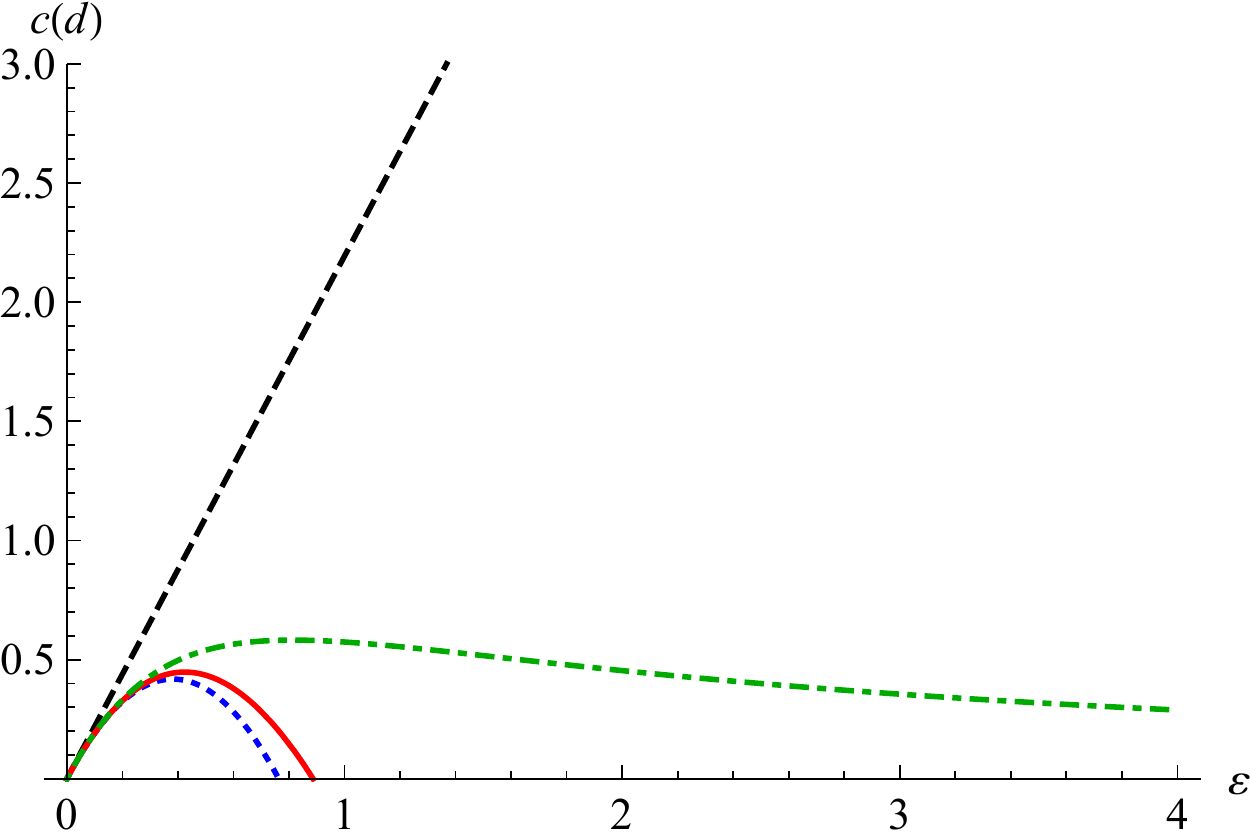}
\caption{Left: Dimensional dependence of the universal amplitude $\tilde c(d)$ in the periodic case. Comparison of 1-loop (dashed black), 2-loop (blue dotted), and 3-loop (red line). The green line corresponds to a (1,2)-Pad\'e approximant of the 3-loop solution. Right: ibid for $c(d)$.}
\label{fig:univAmplitudeperiodicfixpoint}
\end{center}
\end{figure}

In order to allow for a periodic solution of the fix-point equation we set $\zeta=0$. Further we assume a period of one; we can  use  the reparametrization invariance in Eq.~(\ref{eq:reparaminv}) to adjust to other periods. The ansatz
\begin{align}\label{9.42}
 \tilde{R}(u) &= (a_1\epsilon +a_2\epsilon^2 +a_3\epsilon^3 +\ldots) +(b_1\epsilon +b_2\epsilon^2 +b_3\epsilon^3 +\ldots)u^2(1-u)^2 +  \ca{O}(\epsilon^4)
\end{align}
works to all  orders in $\epsilon$. This can   be seen from the following observations:
Each further order in a loop-expansion has one more factor of $R(u)$, and 4 more derivatives. So the RG-equations  close for a polynomial up to order $u^{4}$, and  no higher-order terms in $u$ are needed. This leaves us with 5 terms, $u^{i}$, with $0\le i\le 4$.
The function must further be even under the transformation $ u\to 1-u$. This leaves space in \Eq{9.42} for one additional term, $c u(1-u)$, where the constant $c$ may depend on $\epsilon$.  However, each term in the $\beta$-function except the first one $\epsilon R(u)$ has at least two derivatives, so this term would only appear in $\epsilon R(u)$, and thus must vanish. (It can appear at depinning for different reasons, see \cite{LeDoussalWieseChauve2002}.)
This leads to the fix-point function
\begin{align}\label{RP-fixed-point}
\tilde{R}^*(u)&= \frac{\epsilon}{2592} + \frac{\epsilon^2}{7776} +
 \epsilon^3 \left(-\frac{1}{46656} + \frac{\pi^2}{23328} - \frac{\psi'( \sfrac{1}{3})}{15552} +
    \frac{\zeta(3)}{15552}\right) \\ &\nonumber \quad - (1 - u)^2 u^2 \left(\frac{\epsilon}{72} + \frac{\epsilon^2}{108}+
    \epsilon^3 \frac{9 + 2 \pi^2 - 3 \psi'( \sfrac{1}{3}) - 18 \zeta(3)}{1944}\right)   + \ca{O}(\epsilon^4)\ .
\end{align}
With numerical coefficients, the function reads
\begin{align}
\tilde{R}(u)
&\approx 0.000385802 \epsilon + 0.000128601 \epsilon^2 -
 0.000170212 \epsilon^3 \nn \\ & \quad - (0.0138889 \epsilon + 0.00925926 \epsilon^2 -
    0.0119262 \epsilon^3) (1- u)^2 u^2 + \ca{O}(\epsilon^4) \ .
\end{align}
Similarly as for random-{field} disorder we obtain the universal amplitude as
\begin{align}
 \tilde{c}(d)=-\frac{1}{\epsilon \tilde{I}_1} \tilde{R}''(0)  \approx 2.19325 \epsilon - 0.680427 \epsilon^2 - 2.71612 \epsilon^3 + \ca{O}(\epsilon^4)\ .
\end{align}
This is the 2-point correlation function at zero momentum. There is a large contribution in 3-loop order with a larger coefficient than at 2-loop order. For $\epsilon>0.72$ the 3-loop expansion becomes   negative (as does the 2-loop expansion for $\epsilon>3.22$). This makes the $\epsilon$-expansion questionable in this case, although the (1,2)-Pad\'e approximant remains positive,
\begin{align}
\tilde{c}(d)_{(1,2)} &\approx\frac{2.19325 \epsilon}{1 + 0.310238 \epsilon + 1.33465 \epsilon^2}+ \ca{O}(\epsilon^4)\ .
 \end{align}
The results from different truncations in the loop order and the (1,2)-Pad\'e approximant are plotted in Fig.~\ref{fig:univAmplitudeperiodicfixpoint}. The amplitude of the propagator in the massless limit is given by
\begin{align}
 c(d) \approx 2.19325 \epsilon - 2.87367 \epsilon^2  + 0.45 (1) \epsilon^3+\ca{O}(\epsilon^4)\ ,
\end{align}
with a 3-loop coefficient not as large as the 2-loop coefficient. Here, however, already the 2-loop solution leads to negative values for $\epsilon> 0.76$. The probably best extrapolations  is obtained from the  (1,2)-Pad\'e approximant
\begin{equation}
 c(d)_{{(1,2)}}  \approx\frac{2.19325 \epsilon }{1+ 1.31024\epsilon +1.510(6) \epsilon ^2}+ \ca{O}(\epsilon^4) \ .
\end{equation}

\section{The correction-to-scaling exponent $\omega$}
The correction-to-scaling exponent $\omega$ controls what happens when a fixed point, here a functional fixed point, is  perturbed.  In particular, for a fixed point $\Delta^*(u)=-\partial_u^2{{R}^*}(u)$ with $\beta[\Delta^*]=0$ we consider linear perturbations. Their eigenvalue $\omega$ is determined from the $\cal O(\kappa)$-term in the equation
\begin{equation}
\beta[\Delta^*+\kappa z](u) = \omega \kappa \; z(u) +\ca O(\kappa^2)\, .
\end{equation}
Since observables, and also scaling functions which determine the critical exponents, in general depend analytically on the coupling constants, a deviation of a  critical exponent from the fix-point value scales linearly with the deviation of the coupling constant, or coupling function, from its value at the critical point. In formulas, an observable $\cal O$ or exponent $\alpha$ scales with a length scale $\ell$ as
\begin{equation}\label{10.2}
\ca O -\ca O_{\rm fix-point}\sim
\alpha-\alpha_{\rm fix-point} \sim R(u)- R_{\rm fix-point}(u)\sim \ell^{-\omega}\ .
\end{equation}
This is important for numerical simulations, where $\ell$ is the system size.

For disordered elastic manifolds, this problem has been considered in Ref.~\cite{LeDoussalWiese2003a}. There it was concluded, that two cases have to be distinguished:
\begin{itemize}
\item [(a)] There is the freedom to rescale
the field $u$ while at the same time rescaling the disorder
correlator. This includes the random-bond and random-field interface
models.
\item [(b)] There is no such freedom, since the period is fixed by the
microscopic disorder. This is the case for a charge density wave
(random periodic problem), but also for the random-field bulk problem,
in its treatment via a non-linear sigma model.
\end{itemize}
In case (a), the two leading eigenvalues and eigenfunctions
to linear order in $\epsilon$ are \cite{LeDoussalWiese2003a}
\begin{eqnarray}\label{t2.0}
z_{\rm red } (u) &=& u \Delta' (u) -2 \Delta (u)\ , \qquad ~~~~~~~~~~~~~\omega_{\rm red} = 0 \ , \label{lf70.0a}\\
\label{t2}
z_{1} (u) &=& \zeta u \Delta' (u) + (\epsilon -2\zeta) \Delta (u) \ , \qquad
\omega_{1} = -\epsilon \ . \label{lf70.a}
\end{eqnarray}
The first eigenvalue and eigenfunction $z_{\rm red } (u)$ are a consequence of the reparametrization invariance $\Delta(u)\to \kappa^2 \Delta(u/\kappa)$, and are therefore exact. $z_{\rm red } (u)$ is a {\em redundant} operator.
$z_1(u)$ and $\omega_1$ are the dominant eigenfunction and eigenvalue entering into \Eq{10.2}.
Both eigenfunctions are given as perturbations of the fixed point $\Delta(u)$ of the force-force correlator.
At least for random-field disorder, is was   argued \cite{LeDoussalWiese2003a} that there cannot be any other eigenvalues and eigenfunctions.

In case (b) we can at 1-loop order identify two perturbations, written here as perturbations for the potential-potential correlator $R(u)$:
\begin{eqnarray}
  z_{0} (u)&=& 1\ , \qquad ~~~~~~\omega_{0}=\epsilon\ ,\\
 z_1 (u)&=& R(u)\ , \qquad \omega_1 =-\epsilon\ .
\end{eqnarray}

\subsection{The correction-to-scaling exponent $\omega$ to 2-loop order: General formulas}
The 2-loop $\beta$-function is
\begin{equation}
-m\partial_{m}\tilde \Delta(u) = (\epsilon -2 \zeta )\Delta (u) +\zeta  u \Delta '(u) + f_1[\Delta,\Delta](u) +f_2[\Delta,\Delta,\Delta](u)+... \label{31}
\end{equation}
Both $f_{1}[\Delta]\equiv f_{1}[\Delta,\Delta]$ and $f_{2}[\Delta] \equiv f_{2}[\Delta,\Delta,\Delta]$ are completely symmetric functionals acting locally on the functions $\Delta(u)-\Delta(0)$. More explicitly, we have
\begin{equation}
f_{1}[\Delta]=- \frac{1}{2} \left[(\Delta(u) - \Delta(0))^2\right]''\ ,
\end{equation}
\begin{eqnarray}
f_{2}[\Delta]&=&\frac{1}{2} \left[ (  \Delta(u) -   \Delta(0))  
\Delta'(u)^2 \right]'' - \frac{1}{2}   \Delta'(0^+)^2   \Delta''(u)\ .
\end{eqnarray}
For different arguments we use the multilinear formulas
\begin{eqnarray}\label{lf2}
f(x,y) &:=& \frac12 \Big[f(x+y)-f(x)-f(y)\Big] \ , \\
g (x,y,z) &:=& \frac{1}{6}\Big[ g (x+y+z) - g (x+y)- g (y+z)- g (x+z)+ g (x)+ g (y)+ g (z) \Big] \ ,  \\
h (w,x,y,z) &:=&\frac{1}{24}\Big[h(w+x+y+z)-h(w+x+y)-h(w+x+z)-h(w+y+z)\nn \\
&&\qquad -h(x+y+z) +h(w+x) +h(w+y)+h(w+z)+h(x+y)+h(x+z)\nn \\
&& \qquad +h(y+z)-h(w)-h(x)-h(y)-h(z)\Big]\ .
\end{eqnarray}
Consider now $\tilde   \Delta^*(u)$,   solution of Eq.~(\ref{31})  with $-m\partial_{m}\tilde \Delta^*(u)=0$.
Setting  $\tilde \Delta(u) = \tilde \Delta^*(u) + \kappa z(u)$, we   study the flow of the term linear  in $\kappa$. Its eigenmodes $z(u)$ with   eigenvalues $\omega$ describe  the behavior close to the critical point.
The eigenvalue-equation to be solved is
\begin{eqnarray}
o(u)&:=&\big[\epsilon -2 \zeta  -\omega \big] z(u) + \zeta    u z'(u)+ 2 f_1[z,\Delta](u)+3 f_2[z,\Delta,\Delta] (u)=0\ .~~~~~
\label{36}
\end{eqnarray}
There are several possible simplifications. First note that  if $\Delta(u)$ is a fixed point, also $\kappa^{-2}\Delta(\kappa u)$ is a fixed point. Varying in Eq.~(\ref{31}) the fixed-point condition $-m\partial_m\Delta(u)=0$ around $\kappa=1$ yields the {\em redundant } or {\em rescaling mode} $r(u)$,
\begin{eqnarray}\label{t4-2}
r(u)&=&  (\epsilon
-2 \zeta) \left(u \Delta' (u)-2 \Delta (u) \right) + \zeta u \left(u
\Delta' (u)-2 \Delta (u) \right)' \nonumber \\
& & + 2 f_{1} \left( \Delta (u),
u \Delta' (u)-2 \Delta (u) \right)  + 3 f_{2} \left( \Delta(u), \Delta(u),
u \Delta' (u)-2 \Delta (u) \right)  =0 \ .\ \ ~~
\end{eqnarray}
This equation, as well as a multiple of the vanishing $\beta$-function  (\ref{31}), can be added to Eq.~(\ref{36}). This leaves  some freedom to obtain  a simpler equation.

We now want to know how the physically relevant correction-to-scaling exponent $\omega=-\epsilon$ changes to 2-loop order. To this aim we do a loop expansion, starting from what we know,
\begin{eqnarray}
\Delta(u)&=&\epsilon \Delta_{1}(u)+\epsilon^{2 } \Delta_{2}(u)+...\\
z(u)&=& \epsilon z_{1}(u)+\epsilon^{2} z_{2}(u)+...\\
z_{1} (u) &=& \zeta_{1} u \Delta_{1}' (u) + (1 -2\zeta_{1}) \Delta_{1} (u) \\
\zeta&=&\zeta_{1}\epsilon+\zeta_{2}\epsilon^{2}+...\\
\omega&=&- \epsilon + \omega_{2}\epsilon^{2}+...
\end{eqnarray}
The 1- and 2-loop orders of the $\beta$-function are given by $\beta = \epsilon \beta_1 +\epsilon^2\beta_2 + \ca{O}(\epsilon^3)$ with
\begin{eqnarray}
\beta_{1}&=&(1-2 \zeta_{1})\Delta _1(u)+\zeta _1 u \Delta
   _1'(u)+f_1\left(\Delta _1\right)(u)=0\ ,  \\
\beta_{2}&=& (1-2
   \zeta _1  )\Delta _2(u)+\zeta _1 u \Delta _2'(u) -2 \zeta _2 \Delta _1(u)+\zeta _2 u \Delta
   _1'(u)+2 f_1(\Delta _1,\Delta _2)+f_2(\Delta
   _1)(u)\ .~~~~~~~~~~~\end{eqnarray}
There are many ways a relatively simple differential relation for $\delta z_{2}(u)$ can be written.
We start with the ansatz
\begin{equation}
z_{2}(u) = c u \Delta _1'(u)+d \Delta _1(u)+e u \Delta _2'(u)+f \Delta
   _2(u)+\delta z_2(u)\ ,
\end{equation}
and consider the following combination
\begin{equation}
o(u)-\beta (u) (2+ \epsilon  (4 b+2 d))-r(u) (\zeta_{1}+b \epsilon )-g \beta_{2}(u) \epsilon^{3} =0\ .
\end{equation}
For \beq
b=c=\frac{2 \zeta
   _2}{1-2 \zeta _1} \ , \quad
d= -\omega_2\ , \quad
e= \zeta _1\ , \quad
f= 2-2 \zeta _1\ , \quad
g=1\ ,
\eeq
we get
\beq
2\left(1-\zeta _1\right) \delta z_2(u)+\zeta _1 u \delta z_2'(u)
+
2 f_1\left[\Delta _1(u),\delta z_2(u)\right] -\omega _2
   \Delta _1(u)+\Delta _2(u)=0\ .
\eeq
This is the simplest equation we have been able to find.

At 3-loop order, the problem becomes more complicated. The best equation we found was
\begin{eqnarray}
&& 2 f_1\Big(\Delta _1(u),\delta z_3(u)\Big)+2 f_1\Big(\Delta _2(u),\delta
   z_2(u)\Big)+3 f_2\Big(\Delta _1(u),\Delta _1(u),\delta z_2(u)\Big)\nn\\
   && +\left(\frac{4
   \zeta _2 \omega _2}{1-2 \zeta _1}-\omega _2^2+\omega _3\right) f_1\Big(\Delta _1(u),\Delta
   _1(u)\Big)+\left(\frac{4 \zeta _2}{1-2 \zeta _1}-3 \omega _2\right) \Delta _2(u)+2 \Delta
   _3(u)\nn\\
   && -\left(2 \zeta _2+\omega _2\right) \delta z_2(u)+\zeta _2 u \delta z_2'(u)+\zeta _1 u \delta z_3'(u)-2 (\zeta _1-1) \delta z_3(u) = 0\ .
\end{eqnarray}
For the lack of use in  applications (the 3-loop order for the roughness exponent is rather large), we did not try to solve this equation.

We now specify to the  main cases of interest.
\subsection{Correction-to-scaling exponent at the random-field fixed point}
\begin{figure}[t]
\pfig{5.5cm}{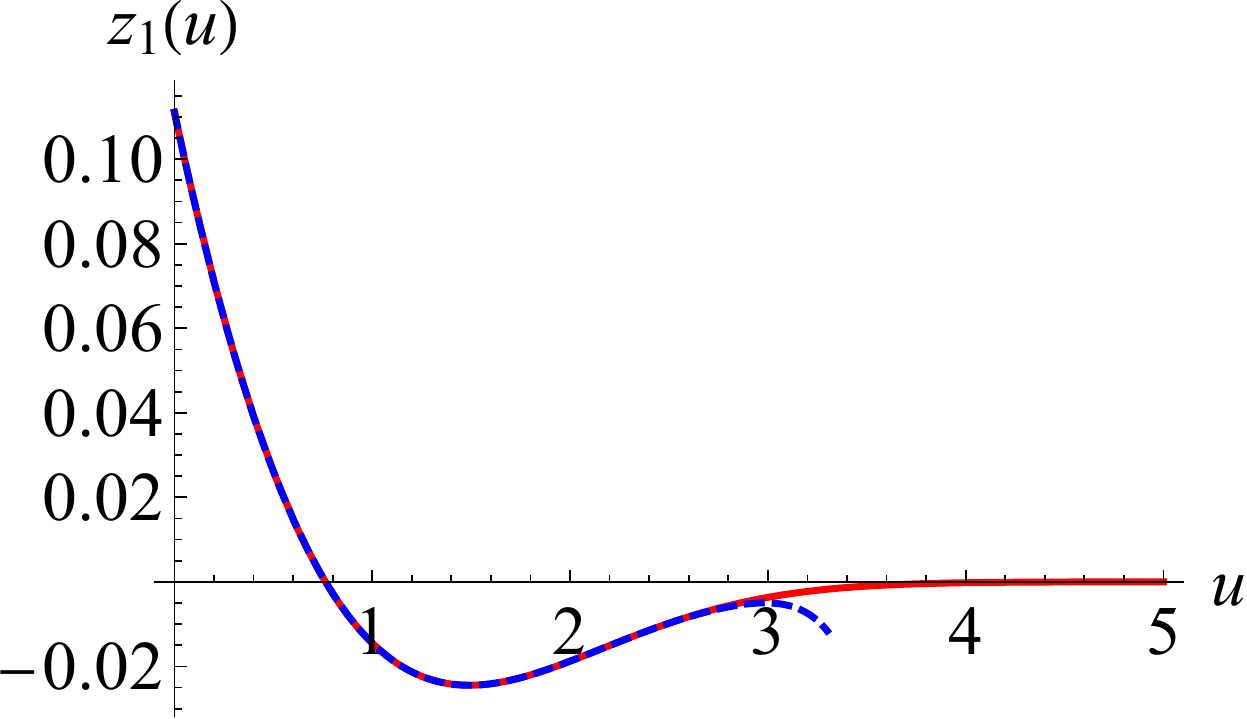}~~\pfig{5.5cm}{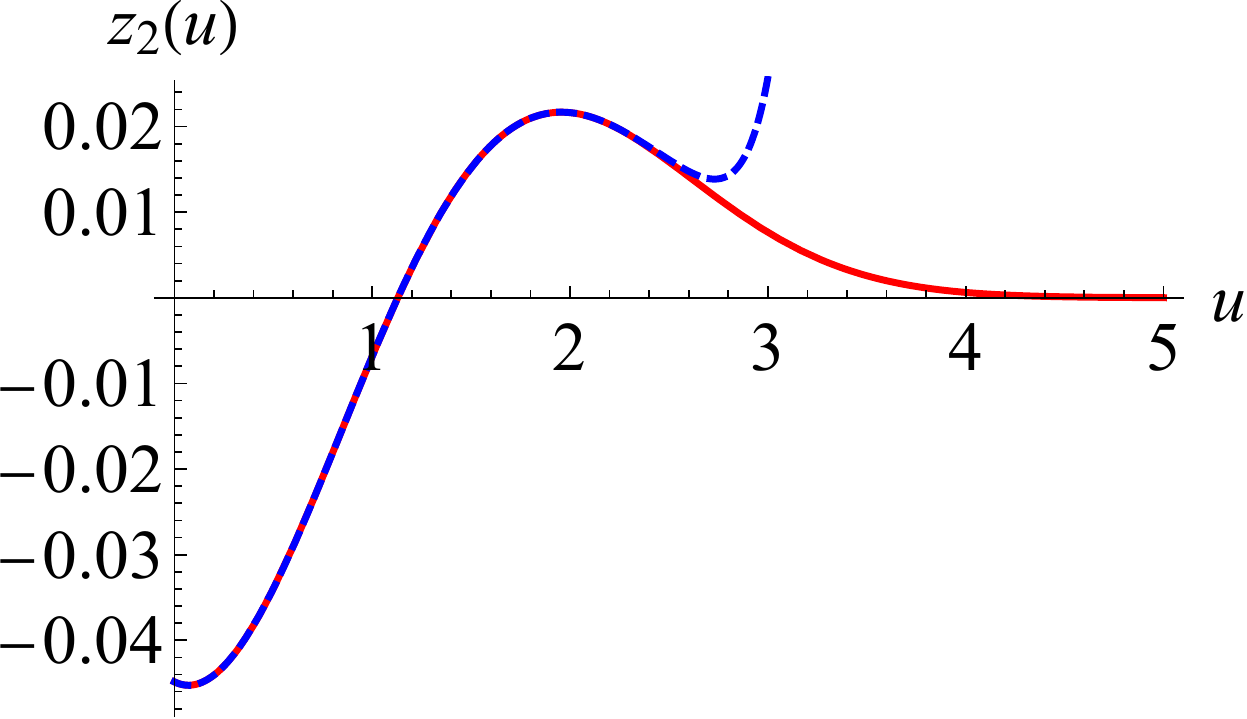}~~\pfig{5.5cm}{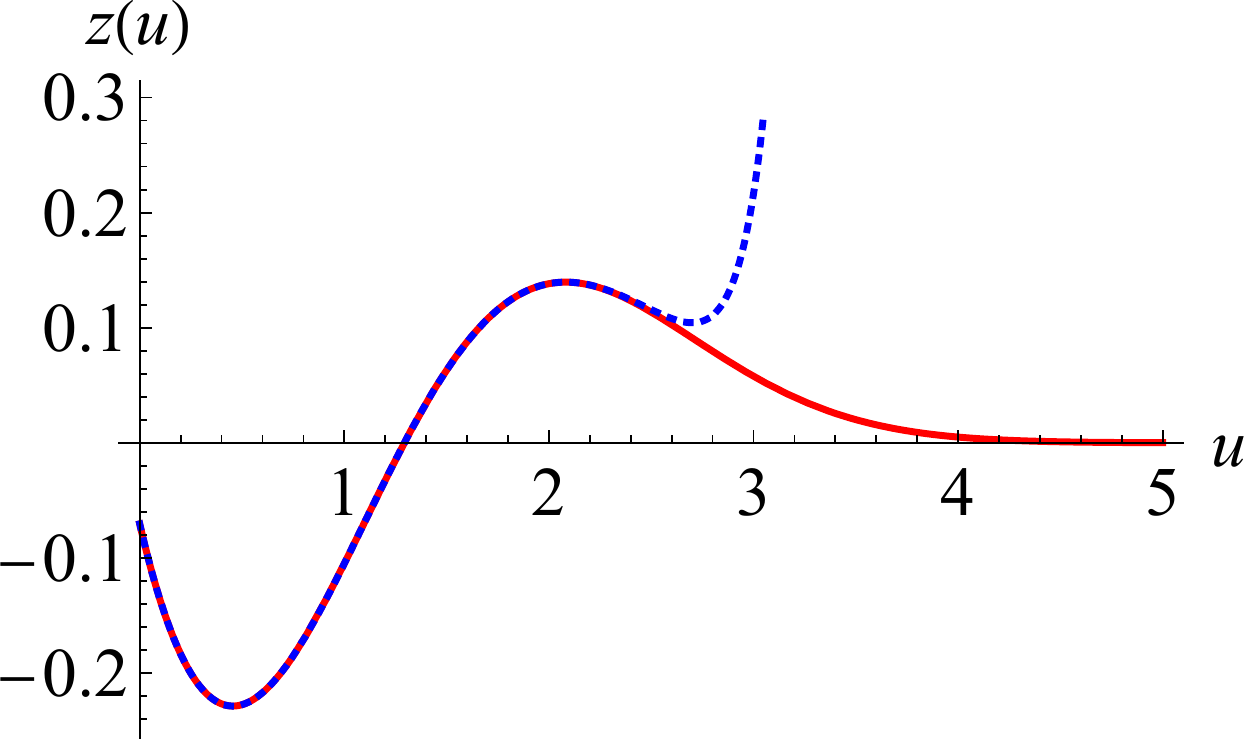}
\caption{$z_{1}(u)$, $z_{2}(u)$ and $z(u)_{\epsilon=3}$ for RF, statics.  Red (solid) is the numerical solution, blue (dashed) the Taylor expansion around $u=0$.}
\label{f:deltazRFstat}
\end{figure}
Using shooting,  we find $
\omega_{2} = 0.1346$, thus
\begin{equation}
\omega \approx -\epsilon +  0.1346 \epsilon ^{2} +{\cal O}(\epsilon^{3})= - \frac{\epsilon}{1+ 0.1346 \epsilon} +{\cal O}(\epsilon^{3})\ .
\end{equation}
The corresponding function $\delta z_{2}(u)$ and $z(u)$ at $\epsilon=3$ are plotted on figure \ref{f:deltazRFstat}. In $d=1$ this gives
\begin{equation}
\omega = -1.97(20)\ ,\qquad d=1\ .
\end{equation} where the error-estimate comes from the deviation of the direct expansion as compared to the  Pad\'e approximant.

\subsection{Correction-to-scaling exponent at the random-bond fixed point}
\begin{figure}[t]
\pfig{5.5cm}{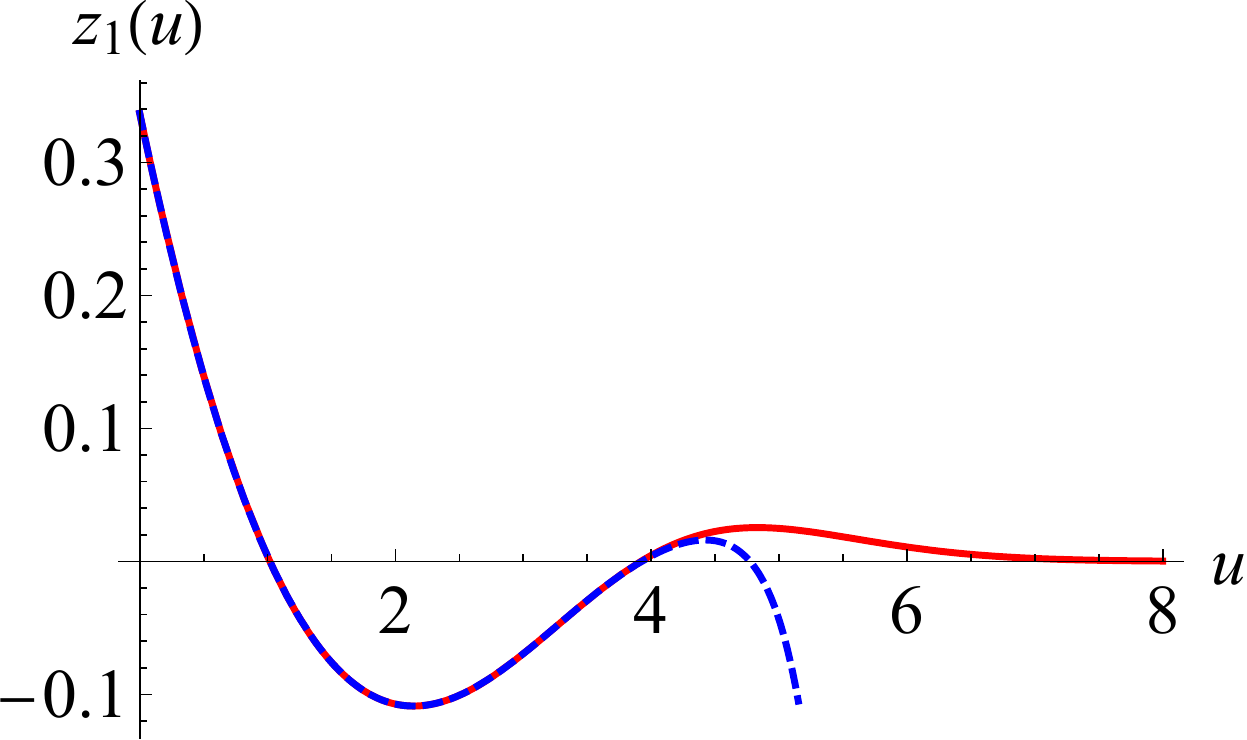}~~\pfig{5.5cm}{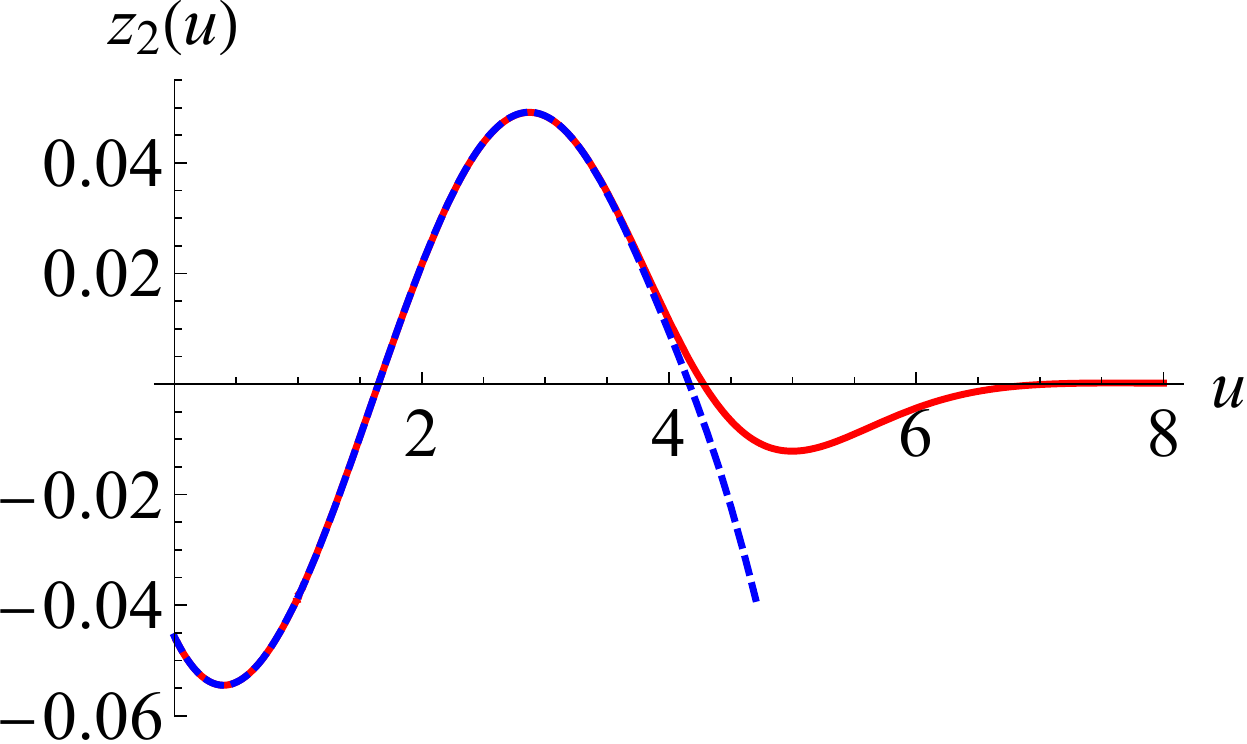}~~\pfig{5.5cm}{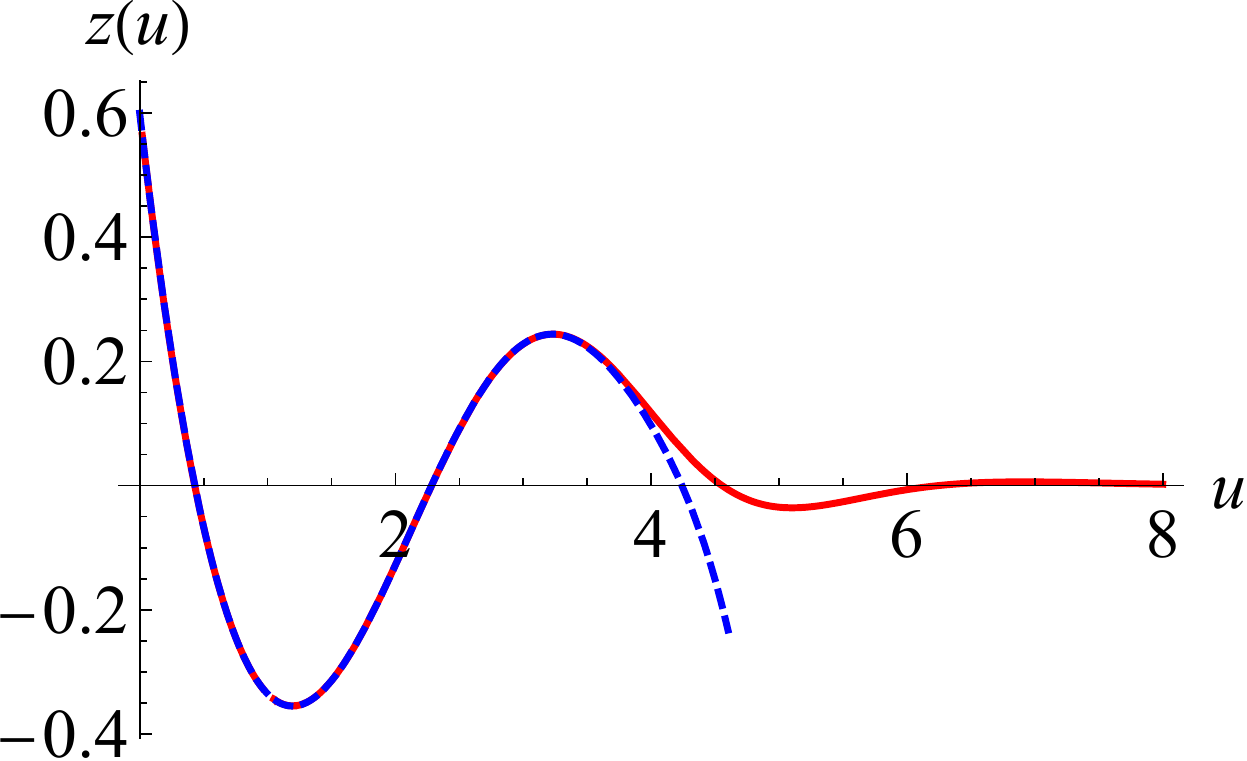}
\caption{$z_{1}(u)$, $z_{2}(u)$ and $z(u)|_{\epsilon=3}$ for RB, statics.  Red (solid) is the numerical solution, blue (dashed) Taylor expansion.}
\label{f:deltazRB}
\end{figure}

We find via shooting \begin{equation}
\omega \approx -\epsilon +0.4108(1) \epsilon^{2} +{\cal O}(\epsilon^{3})= -\frac{\epsilon }{1+0.4108(1) \epsilon }+{\cal O}(\epsilon^{3})\ .
\end{equation}
In $d=1$ this gives using the Pad\'e approximant (the direct $\epsilon$ expansion is not monotonous)
\begin{equation}
\omega  \approx -1.344\ ,\qquad d=1\ .
\end{equation}
We have checked that the numerical solutions, given on figure \ref{f:deltazRB}, integrate to 0 within numerical accuracy, as necessary for a RB fixed point.

\subsection{Correction-to-scaling exponent for charge-density waves (random-periodic fixed point)}
We find that the leading-order perturbation for the random-periodic fix-point (\ref{RP-fixed-point}) closes in the same space spanned by 1 and $[u(1-u)]^2$. The correction-to-scaling exponent becomes
\begin{equation}
\omega_{\rm RP} = -\epsilon +\frac{2 \epsilon ^2}{3}-\left(\frac{4 \zeta
   (3)}{3}+\frac{5}{9}\right) \epsilon ^3     +{\cal O}(\epsilon^{4})= -\epsilon\, \frac{1 + \left[2 \zeta (3)+\frac{1}{6}\right] \epsilon }{1+\left[ 2 \zeta (3)+\frac{5}{6}\right]
   \epsilon } +{\cal O}(\epsilon^{4})\ .
\end{equation}
Curiously, all contributions proportional to $\pi^2$ and $\psi'(1/3)$, present in the coefficients $\ca C_1,...,\ca C_4$ have canceled.
The corresponding eigenfunction, normalized to $\delta R(0)=1$ is
\beq
\delta R(u) = 1 - 36 [u(1-u)]^2 \left[ 1+\frac \epsilon2 +\epsilon^2 \frac{8 -30 \zeta (3)+3 \psi ' (\frac{1}{3}   )-2 \pi ^2}{18}   + \ca O(\epsilon ^3)\right]\ .
\eeq
Up to 2-loop order, the exponent $\omega$ is the same for depinning.

\subsection{Correction-to-scaling exponent for depinning (random-field fixed point)}
\begin{figure}
\pfig{5.5cm}{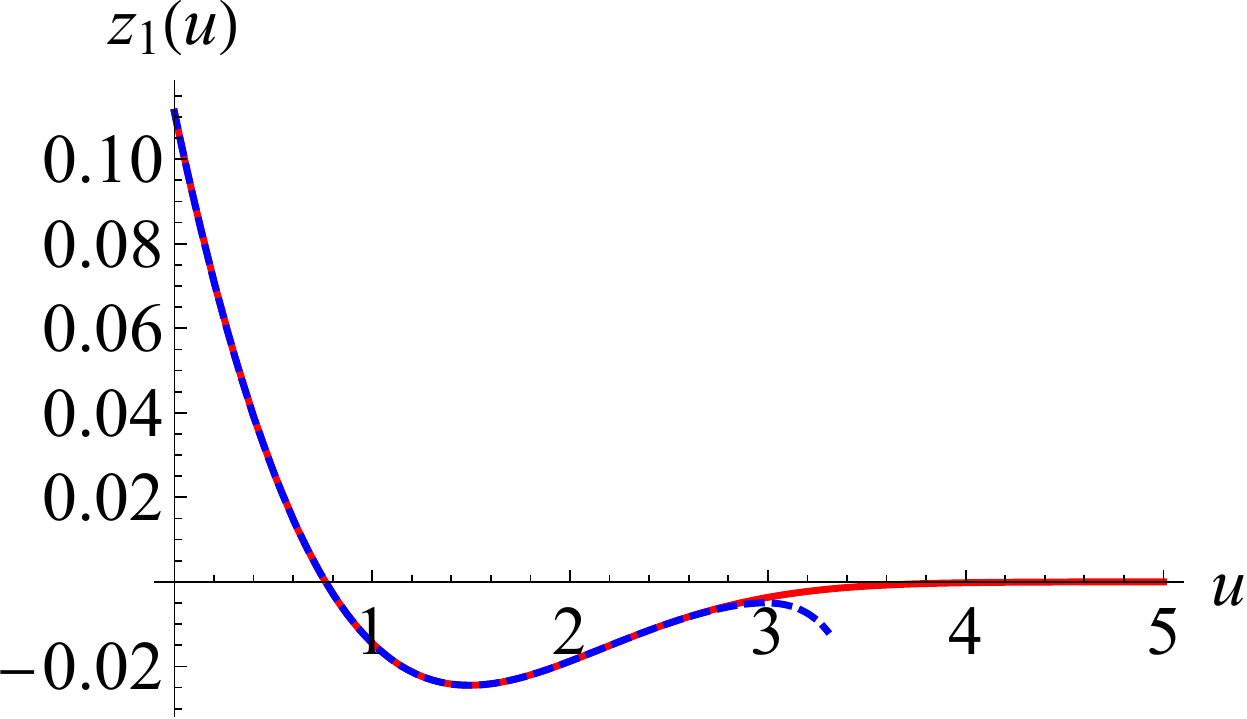}~~\pfig{5.5cm}{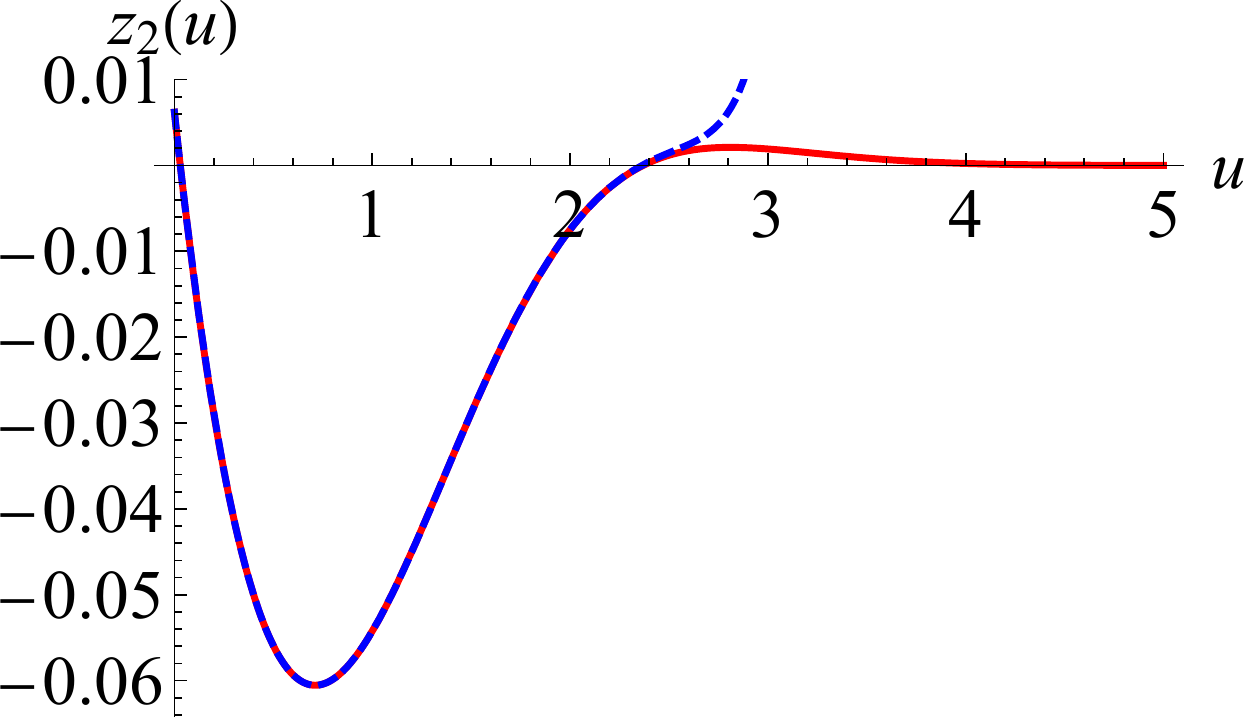}~~\pfig{5.5cm}{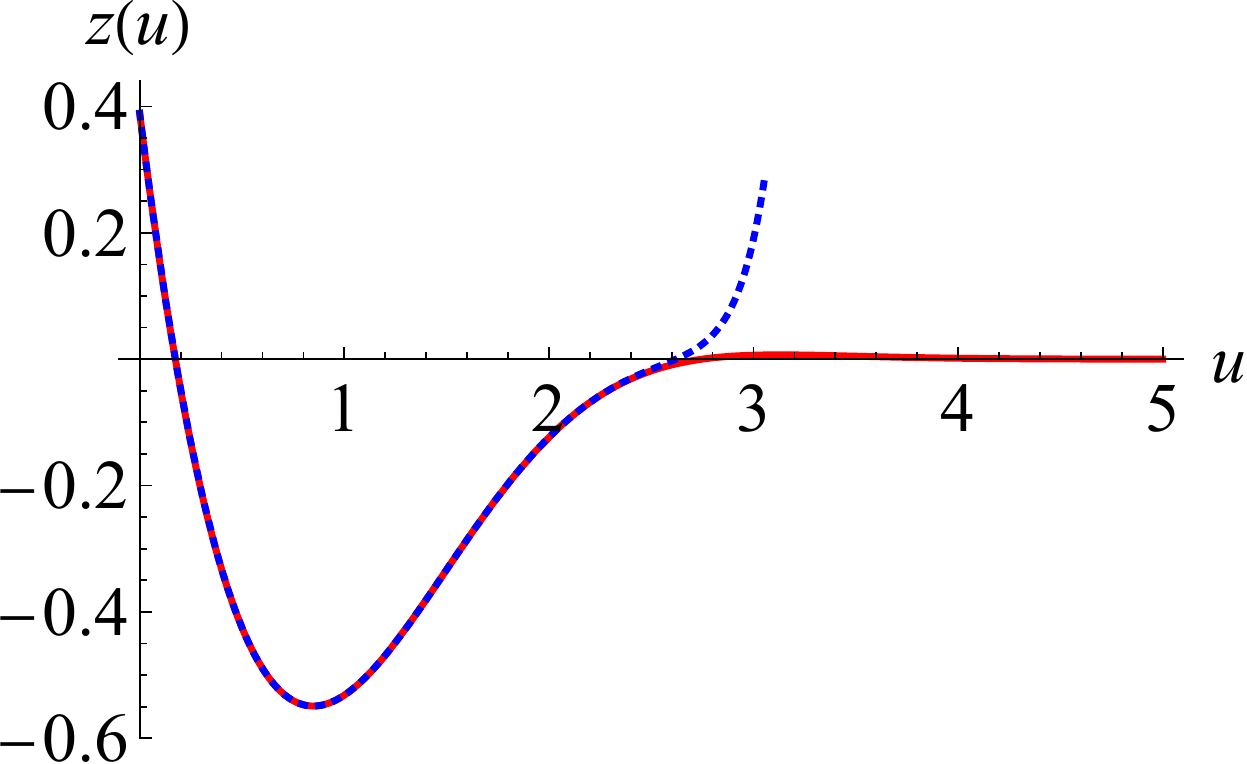}
\caption{$z_{1}(u)$, $z_{2}(u)$ and $z(u)|_{\epsilon=3}$ for RF, depinning. Red (solid) is the numerical solution, blue (dashed) the Taylor expansion.}
\label{f:deltazRFdep}
\end{figure}

For completeness and usefulness in applications, we also give the correction-to-scaling exponent at depinning, using the $\beta$-function of \cite{LeDoussalWieseChauve2002,ChauveLeDoussalWiese2000a}.
Via shooting, we find $\omega_{2} = -0.0186$, thus
\begin{equation}
\omega\approx - \epsilon  - 0.0186\epsilon^{2} + {\cal O}(\epsilon^{3})
= -\frac{\epsilon }{1-  0.0186\epsilon} + {\cal O}(\epsilon^{3})\ .
\end{equation}
The corresponding function $\delta z_{2}(u)$ and $z(u)$ at $\epsilon=3$ are plotted on figure \ref{f:deltazRFdep}. In $d=1$ this gives
\begin{equation}
\omega = -3.17(1)\ , \qquad d=1\ .
\end{equation} where the error-estimate comes from the difference of the Pad\'e approximant to the direct expansion.

\section{2-point correlation function}
\subsection{2-loop expression}
As a prototype physical observable we calculate the 2-point correlation function to 2-loop order in an $\epsilon$-expansion. This function  reads
\begin{align}\label{eq:TwoPointCorrelationRealSpace}
 \ovl{\langle u(x)u(y)\rangle}_V = T g(x,y) -{ \int_{z,z'} g(x,z)g(y,z')} R''[0^+](z,z') \, .
\end{align}
The expression $R''[0^+](z,z')$ denotes the second functional derivative of $R[u]$ with respect to $u(z)$ and $u(z')$ that is evaluated for $u(x)= \mbox{const} \to 0$. Its expansion to 2-loop order has been calculated in \cite{WieseHusemannLeDoussal2018} and reads
\begin{align}
R''[0^+](z,z')= R''(0^+)\delta(z-z') + \tilde{R}''[0^+](z,z') \, ,
\end{align}
where the non-local part is given by
\begin{align}
\tilde{R}''[0^+](z_1,z_2) &= \delta_{z_1z_2} \Big[-I_1 R'''(0^+)^2 +(5I_1^2-4I_A) R'''(0^+)^2 R''''(0^+) \Big] \\
&+ g(z_1,z_2)^2 \Big[R'''(0^+)^2 -6I_1R'''(0^+)^2 R''''(0^+) \Big] \nonumber\\
&+ 2g(z_1,z_2) R'''(0^+)^2 R''''(0^+) \int_x \; g(x,z_1)g(x,z_2)\big[g(x,z_1)+g(x,z_2)\big] \nonumber \\ &+ R'''(0^+)^2 R''''(0^+) \int _x \;g(x,z_1)^2 g(x,z_2)^2 \nonumber .
\end{align}
Inserting into Eq.~(\ref{eq:TwoPointCorrelationRealSpace}) and taking the Fourier transform gives at $T=0$
\begin{align}\label{eq:TwoPointTwoLoop}
g(q)^{-2} \ovl{\langle u(q)u(-q)\rangle}_V &= -R''(0^+) +R'''(0^+)^2 \big[ I_1-I_1(q) \big] \\ &\quad - R'''(0^+)^2 R''''(0^+) \big[ (I_1-I_1(q))^2 + 4 {\Phi_{2,\epsilon}}(q)  \big]\ . \nonumber
\end{align}
As before $I_1=I_1(0)$, $I_A=I_A(0)$, and $\Phi_{2,\epsilon}(q)= I_A(q)-I_A + I_1^2 - I_1I_1(q)$ with
\begin{align}
I_1(q) &=\; \stackrel q{\longrightarrow}\!\!\!-\!\!\!-\!\!\!\!\diagram{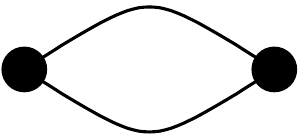}\!\!\!\!\!\!\longrightarrow\!\!\!-\!\!\!- = \int _p \; g(p) g(p+q) \\
I_A(q) &=\; \raisebox{-3.6mm}{$\stackrel q{\longrightarrow}\!\!\!-\!\!\!-\!\!\!$}\diagram{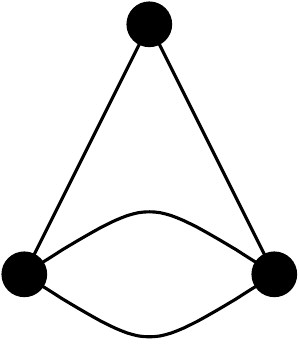}\!\!\!\!\!\!\!\!\!\raisebox{5.3mm}{${\longrightarrow}\!\!\!-\!\!\!-\!\!\!$} =\int _{p_1,p_2}g(p_1)g(p_2) g(p_1+p_2) g(p_1+q) .\label{8.6}
\end{align}
The first line in Eq.~(\ref{eq:TwoPointTwoLoop}) is the tree and 1-loop contribution, already calculated in \cite{LeDoussalWieseChauve2003}. The second line is the 2-loop contribution, again with a non-vanishing  limit  $R'''(0^+)^2$. Note that the momentum could go through the non-trivial 2-loop diagram  $I_A$ in two different ways, but only the one in Eq.~(\ref{8.6}) contributes to the 2-point function.

\subsection{Integrals}
Let us start out with the normalization factors used throughout this work. Consider the 1-loop integral
\begin{equation}
\tilde I_1 = \diagram{1loop}\Big|_{m=1} = \int _k \; {  g(k)^2 }\Big|_{m=1} =\int \frac{\rmd ^d k}{(2\pi )^d} \frac{1}{(k^2+1)^2} \ .
\end{equation}
Using a Feynman-representation for the propagator yields
\beq
\tilde I_1 = \int \frac{\rmd ^d k}{(2\pi )^d}  \int_0^\infty \rmd \alpha \, \alpha \, \rme^{-\alpha(k^2+1)}
= \int \frac{\rmd ^d k}{(2\pi )^d}  \rme^{-  k^2} \times \int_0^\infty \rmd \alpha \, \alpha^{1-d/2}  \, \rme^{-\alpha}\ .
\eeq
Let us note, with $d=4-\varepsilon$
\begin{align}\label{N}
&{\cal N} :=  \int \frac{\rmd ^d k}{(2\pi )^d}  \rme^{-  k^2} = \frac1{(4\pi )^{\frac d2}}\ , \\
& \int_0^\infty \rmd \alpha \, \alpha^{1-d/2}  \, \rme^{-\alpha} =\Gamma(2-\tfrac d2) = \Gamma(\tfrac \epsilon 2)= \frac2\epsilon \Gamma(1+\tfrac \epsilon2)\ .
\label{8.10}
\end{align}
This yields
\begin{align} \label{epsI1}
& \epsilon\tilde I_1 = {\cal N} \times 2  \Gamma(1+\tfrac \epsilon2) = {\ca N}\times \left[ 2-\gamma_{\rm E}  \epsilon +\frac{1}{24}  (6 \gamma_{\rm E} ^2+\pi ^2 )
   \epsilon ^2+\ca O (\epsilon ^3 )\right] \ .
\end{align}
There are therefore two convenient choices for normalizations:
Either we normalize everything by $ \epsilon\tilde I_1$: then $\tilde I_1$ will be effectively $1/\epsilon$, without higher-loop corrections. Or we normalize by $\ca N$, which   takes out the  factor from the Gauss integration.
We use whatever is more convenient.

We now turn to the evaluation of $I_1(q)$ and $I_A(q)$.
The 1-loop integral in presence of an external momentum can be parameterized by $I_1(q)=\frac{1}{m^{\epsilon}}\tilde{I}_1(\sfrac{q}{m})$ with
\begin{align}
\tilde I_1(z) &= \tilde I_1(0) \; {}_2F_1(2-d/2,1,3/2,-z^2/4)
\label{9.7} \\
&=   {\tilde I_1(0) \int\limits_0^1 \rmd y \; \left[1+y(1-y)z^2\right]^{-\frac{\epsilon}{2}} \ .} \label{9.8}
\end{align}
Using the series expansion of the hypergeometric function in Eq.~(\ref{9.7}) or an expansion in $\epsilon$ of the integrand in Eq.~(\ref{9.8}), we find
\begin{eqnarray}
\tilde I_1(z) &=& \tilde{I}_1(0)\Bigg\{  1+ \epsilon \left(1  - \frac{\sqrt{4+z^2}}{z}\mbox{asinh}(z/2) \right) \nn\\
&+&  \epsilon ^2 \Bigg[ 1+ \frac{ \sqrt{ 4+{z^2}}}{ 4 z} \Bigg(  \text{Li}_2\! \Big(\tfrac{1}{2}-\tfrac{z}{2
   \sqrt{ {z^2}+{4} }}\Big)-\text{Li}_2\!\Big(\tfrac{1}{2} +
 \tfrac{z}{2
   \sqrt{ {z^2}{+}{4}}} \Big) +\left(\ln
    (z^2{+}4 ){-}4\right) \mbox{asinh}(z/2) \Bigg)\Bigg] \nn\\
& + &\ca O(\epsilon^3) \Bigg\}\ .
\end{eqnarray}
All singular contributions of $\tilde I_1(z)$ are present for $z=0$, where $\tilde{I}_1(0)=\ca N[ \sfrac{2}{\epsilon} -\gamma_{\mtin{E}} +\ca{O}(\epsilon)]$. Thus, the difference $I_1-I_1(q)$ is finite in the limit of $\epsilon\to 0$. The asymptotics for large $z$ is given by
\begin{eqnarray}
\tilde{I}_1(z) \simeq  \tilde{I}_1(0)\bigg\{ 1 &+& \epsilon\left [1  - \ln z +\ca{O}(\sfrac{1}{z}\ln
z)\right]+\epsilon^2\left[\frac12 (\ln z)^2 - \ln (z)+1-\frac{\pi ^2}{24} +\ca O(\sfrac 1 z\ln z)\right] \nn\\
&  +&\ca O(\epsilon^3) \bigg\}\ .
\end{eqnarray}
The second term $\Phi_{2,\epsilon}(q) = \frac{1}{m^{2\epsilon}} \tilde\Phi_{2,\epsilon}(\frac{q}{m})$ is more complicated and treated in App.~\ref{app:TwoPointCorrelation}. It also has a finite limit $\epsilon\to 0$, which we can only state as an integral. A Taylor expansion for small $z$ gives
\begin{align}
 \tilde\Phi_{2,0}(z) \approx { \ca N^2} \big(  0.03821 z^2 +0.00169 z^4 - 0.00039 z^6 + 0.00007z^8 +\ldots \big)\ .
\end{align}
For large $z$ we find
\begin{align}
  \tilde{\Phi}_{2,0}(z) \simeq { \ca N^2}\big[   2 (\ln z)^2 -6 \ln z + \alpha_0 + \ca{O}(\sfrac{1}{z}\ln z)\big] \ .
\end{align}
Thus the full 2-loop contribution to the 2-point function has the asymptotic form
\begin{align}\label{8.18}
 \big[\tilde{I}_1(z)-\tilde{I}_1(0)\big]^2 + 4 \tilde{\Phi}_{2,0}(z) \simeq  {\ca N^2} \big[ 12(\ln z)^2 -32 \ln z +4 + { 4\alpha_0} +\ca{O}(\sfrac{1}{z}\ln z)+\ca{O}(\epsilon) \big] \,.
\end{align}
The constant $\alpha_0\approx 6.17$ was calculated numerically.

\subsection{Scaling function (for arbitrary $\zeta$)}
\label{s:scaling-function-for-aribtrary-zeta}
We parameterize the 2-point correlation function as
\begin{align}
{\overline{ \langle   u(q)   u(-q) \rangle} } = m^{-d-2\zeta} \tilde{c}(d) F_d(\sfrac{|q|}{m})
\end{align}
with universal amplitude $\tilde{c}(d)$ and scaling function $F_d$ with $F_d(0)=1$. At momentum zero the higher-loop terms do not contribute to the 2-point correlation function such that to all orders in $\epsilon$
\begin{align}
 \tilde{c}(d) = m^{d+2\zeta} \langle u(0) u(0) \rangle = -\frac{1}{ \epsilon \tilde{I}_1} \tilde{R}''(0) .
\end{align}
Using Eq.~(\ref{eq:TwoPointTwoLoop}) and the rescaled renormalized disorder $\tilde{R}$, defined in Eq.~(\ref{R-rescale}), the scaling function is given by
\begin{align}\label{eq:scalinghelp0}
 F_d(z) &= \frac{ -1}{\tilde{R}''(0)} \frac{1}{(1+z^2)^2} \left\{ -\tilde{R}''(0) +\tilde{R}'''(0^+)^2 \frac{1}{\epsilon \tilde{I}_1} \left[ \tilde{I}_1 -\tilde{I}_1(z)\right] \right. \\ & \left.\nonumber \hspace{2cm} -\tilde{R}'''(0^+)^2 \tilde{R}''''(0^+) \frac{1}{(\epsilon \tilde{I}_1)^{ 2}} \left[\left(\tilde{I}_1 -\tilde{I}_1(z)\right)^{\!\!2} +4 \tilde\Phi_{2,\epsilon}(z) \right] +\ca{O}(\epsilon^4) \right\}.
\end{align}
At a fix-point there are a number of consistency relations for the third and fourth derivatives of the disorder distribution   at $u=0^+$ in an $\epsilon$-expansion. Taking two field-derivatives of the fix-point equation, that is, evaluating $0=\frac{\partial}{\partial m} \tilde{R}''(0^+)$ gives
\begin{align}\label{eq:scalinghelp}
 0 = (\epsilon- 2\zeta) \tilde{R}''(0^+) + \tilde{R}'''(0^+)^2 +2 \tilde{R}'''(0^+)^2 \tilde{R}''''(0^+) +\ca{O}(\epsilon^4).
\end{align}
Similarly, $0=\frac{\partial}{\partial m} \tilde{R}'''(0^+)$ gives the identity
\begin{align}
 0&= R'''(0^+) \big[ -\zeta +\epsilon +3 R''''(0^+) +\ca{O}(\epsilon^2)\big].
\end{align}
Regardless of the sign of the prefactor, the bracket has to vanish. Therefore, evaluation at $u=0^-$ would lead to the same result, namely $\tilde{R}''''(0^+)= -\frac{\epsilon-\zeta}{3} +\ca{O}(\epsilon^2)$, which can be inserted into Eq.~(\ref{eq:scalinghelp}) to obtain
\begin{align}
 \tilde{R}'''(0^+)^2 &= -(\epsilon-2\zeta) \tilde{R}''(0^+) \left[ 1+ \sfrac{2}{3} (\epsilon-\zeta)\right] + \ca{O}(\epsilon^4)  \\
 \tilde{R}'''(0^+)^2\tilde{R}''''(0^+) &= \sfrac{1}{3} (\epsilon-2\zeta)(\epsilon-\zeta)  \tilde{R}''(0^+) + \ca{O}(\epsilon^4).
\end{align}
Substituting these expressions into Eq.~(\ref{eq:scalinghelp0}) gives
\begin{align}\label{Fd(z)-final}
 F_d(z) &=  \frac{{ 1}}{(1+z^2)^2} \left\{ 1+(\epsilon-2\zeta)\big[1+\sfrac{2}{3} (\epsilon-\zeta)\big] \frac{1}{\epsilon \tilde{I}_1} \big( \tilde{I}_1-\tilde{I}_1(z)\big) \right.\\& \nonumber \left. \qquad + \sfrac{1}{3}(\epsilon-2\zeta)(\epsilon-\zeta) \frac{1}{(\epsilon \tilde{I}_1)^2} \left[\left(\tilde{I}_1 -\tilde{I}_1(z)\right)^2 +4 \tilde\Phi_{2,\epsilon}(z) \right]\right\} +\ca{O}(\epsilon^3).
\end{align}
For large $z=\frac{|q|}{m}$ the asymptotic behavior of the scaling function is, with $\alpha_0$ given after \Eq{8.18},
\begin{align}\label{8.27}
 F_d(z) &\simeq \frac{ 1}{(1+z^2)^2} \left\{ 1+(\ln z -1) (\epsilon-2\zeta) + \left(\frac 12 (\ln z)^2 -\ln z \right) (\epsilon-2\zeta)^2 \right.  \\ &\nonumber \qquad \left. +(\epsilon-2\zeta) \left[\left(\frac{\alpha_0 -1}{3} -1 +\frac{\pi^2}{24} \right) \epsilon - \frac{\alpha_0 -1}{3} \zeta \right]\right\}+\ca{O}(\epsilon^3)+\ca{O}(\sfrac{1}{z}\ln z).
\end{align}
Assuming the following behavior for large $z$
\begin{align}\label{9.21}
 F_d(z) \simeq \big[1+b_1\epsilon +b_2\epsilon^2 + \ca{O}(\epsilon^3) \big] z^{\epsilon-2\zeta -4}\ ,
\end{align}
the coefficients $b_{1}$ and $b_{2}$ are
\begin{align}
 b_1&=-(1-2\zeta_1) \ , \\
 b_2&=2\zeta_2 +(1-2\zeta_1) \left[ \frac{\alpha_0-1}{3}(1-\zeta_1) -1 +\frac{\pi^2}{24}\right]\ .
\end{align}
In the massless limit, that is $|q|\gg m$, the amplitude of the 2-point correlation function is given by
\begin{align}
 \langle u(q) u(-q) \rangle \sim |q|^{-(d+2\zeta)} c(d)
\end{align}
with amplitude
\beq
c(d)=[1+b_1\epsilon + b_2\epsilon^2 +\ca{O}(\epsilon^3)] \tilde{c}(d).
\eeq
We also note that transforming to real space we have
\begin{eqnarray}\label{8.33}
\frac12 \overline{\left<[u(x)-u(0)]^2 \right> } &=& c(d) \int\frac{\rmd^d q}{(2\pi)^d}\,{ (1-\rme^{i q x}) }|q|^{-d-2\zeta} \nn \\
&=& \frac{
   -\Gamma (-\zeta) c(d)}{(4\pi)^{\frac d2}\Gamma
   \left(\frac{d}{2}+\zeta \right)} \left(\frac x 2\right)^{\!\!2\zeta}\ .
\end{eqnarray}
This expression breaks down for $\zeta\ge 1$. Having in mind the RF fixed point in $d=1$ which has $\zeta=1$, we consider a finite system of size $L$ and periodic boundary conditions,
\bea
\int\frac{\rmd q}{2\pi}\, (1-\rme^{i q x}) |q|^{-3} &\to& \frac2L \sum_{n=1}^{\infty} \frac{1-\cos(\tfrac{2\pi n x}L)}{ (\tfrac{2\pi n }L)^3}\nn\\
&=&-\frac{L^2}{8 \pi ^3} { \left[\text{Li}_3\!\left(\rme^{-\frac{2 i \pi
   x}{L}}\right)+\text{Li}_3\!\left(\rme^{\frac{2 i \pi
   x}{L}}\right)-2 \zeta_{\rm R} (3)\right]} \nn\\
   &=& \frac{x^2 \left[ 3+2  \log \left(\frac{L}{2 \pi
   x}\right)\right]}{4 \pi }+\ca O(x^3)\ .
\eea
To avoid confusion, we have added an index $\rm R$ to the Riemann $\zeta$-function.
When $\zeta>1$, the correlation function (\ref{8.33}) will grow quadratically with the distance $x$, with an $L$-dependent prefactor scaling as $L^{2\zeta -1}$, even though the Fourier-transform has a pure power-law.
\bea
\int\frac{\rmd q}{2\pi}\, (1-\rme^{i q x}) |q|^{-1-2 \zeta} &\to& \frac{L^{2 \zeta +1} }{  (2 \pi ) ^{2 (\zeta +1)} }
   \left[ 2 \zeta_{\rm R} (2 \zeta +1) - \text{Li}_{2 \zeta +1}\left(e^{-\frac{2 i \pi
   x}{L}}\right) - \text{Li}_{2 \zeta +1}\left(e^{\frac{2
   i \pi  x}{L}}\right)\right] ~~~~~~~~\nn\\
   &=& \frac{L^{2 \zeta -1} }{  (2 \pi ) ^{2  \zeta } }
   \zeta_{\rm R} (2 \zeta - 1) x^2 - \frac{x^{2\zeta} \cos(\pi \zeta) \Gamma(-2 \zeta)}{2\pi^2} + ...
\eea
The best studied example is depinning, where $\zeta = \frac54$ (possibly exactly     \cite{GrassbergerDharMohanty2016}). As an example we mention Fig.~1 of Ref.~\cite{FerreroBustingorryKolton2012}, where one sees that  the structure factor, i.e.~Fourier transform of the 2-point function, is a power-law over almost three decades.

\section{Conclusions and open problems}
In this article, we have obtained the functional renormalization-group flow equations for the equilibrium properties of   elastic manifolds  in quenched disorder up to 3-loop order. This allowed us to obtain several critical exponents, especially the roughness exponent, to 3-loop accuracy, for random-bond, random-field, and periodic disorder.
For an elastic string in a random-bond environment, for which we know the exact value $\zeta=\frac23$, the corrections turn out to be quite large. This suggests that convergence of the $\epsilon$-expansion is plagued by the typical problem of renormalized field theory, namely that the perturbation expansion in the coupling is not convergent, but only Borel-summable.
In $\varphi^4$-theory the physical reason for a only Borel-summable series is that the theory with the opposite sign of the coupling is unstable, thus the perturbative expansion cannot be convergent. For the case at hand, this is not evident: Since averaging over disorder leads to attractive inter-replica interactions, making the latter repulsive should make the problem even better defined: a self-attractive polymer is unstable, whereas a self-repelling one has a well-defined fixed point, the self-avoiding polymer fixed point.
The second point which makes us doubt that the theory is only Borel-summable is that when the interaction behaves as  $\int_x g \varphi^{2 \alpha}(x)$, then the standard instanton analysis yields that  $\left< \exp( -\int_x g \varphi^{2 \alpha}(x) ) \right>  = \sum _{n=0}^\infty \frac{(-g)^n}{n!} \left< \left[ \int_x  \varphi^{2 \alpha}(x)\right]^n \right>  $, with   $\left< \left[ \int_x  \varphi^{2 \alpha}(x)\right]^n \right>  \simeq (n!)^{\alpha}$
 for a total of the $n$-th order term being $ (n!)^{\alpha -1}$. The exponent $\alpha$ in the last formula is extracted from the large $\varphi$ behavior of the interaction. For the problem at hand, $R(u)$ has a Gaussian tail, thus the perturbative expansion should  converge!
 This does however not say anything about the result at a given order, here $n=3$.
 It would be interesting to find an exact solution in some limit, which could shed light on this issue.
In some cases, large $N$ (with $N$ being the number of components) provides such a limit. It has however been shown in \cite{LeDoussalWiese2001,LeDoussalWiese2003b} that the $\beta$-function at leading order in $1/N$ is as obtained in 1-loop order. For the order $1/N$-corrections \cite{LeDoussalWiese2004a}, the same problem appears.

\section*{Acknowledgements}
We acknowledge fruitful discussions with Leon Balents, Dima Feldman, Boris Kastening, Pierre Le Doussal, and Andreas Ludwig.

\appendix

\section{Loop integrals}
\label{a:Diagrams}
\subsection{General formulae,  strategy of calculation, and conventions}
We make use of the Schwinger parameterization
\begin{align}
 \frac{1}{A^n} = \frac{1}{\Gamma(n)} \int_0^\infty \mathrm{d} u \; u^{n-1} e^{-uA}
\end{align}
and the $d$-dimensional momentum integration
\begin{align}
 \int \frac{\rmd^d p}{(2\pi)^d}\rme^{-ap ^2} \equiv \int_p e^{-ap^2}= \frac{1}{a^{d/2}} \int_p \rme^{-p
^2} = \frac{1}{a^{d/2}}   \frac1{(4\pi)^{d/2}}\ .
\end{align}
In order to avoid cumbersome appearances of factors like \( \frac1{(4\pi)^{d/2}}\), we will write explicitly the last integral, and will only calculate ratios compared to the leading 1-loop diagram \(I_1\), given in the next section.

We will frequently use the   decomposition trick
\begin{align}
\frac{1}{k^2+1} = \frac{1}{k^2} -\frac{1}{k^2(k^2+1)}\ ,
\end{align}
which works well for dimension $d\le 4$. The reason for the utility of this decomposition is that it allows one to replace the massive propagator by a massless one, which is easier to integrate over, and a term converging faster for large $k$, which finally renders the integration finite.

Special functions which appear are \begin{eqnarray}
\psi(x) &:=& \frac{\Gamma'(x)}{\Gamma(x)} \ ,\\
\psi'(x)&=& \frac{\rmd}{\rmd x}\psi(x)\ .
\end{eqnarray}

\subsection{The 1-loop integral $I_{1}$}\label{app:I1}
The integral $I_{1}$ is defined as
\begin{equation}\label{I1}
I_{1}:= \diagram{1loop}= \int_{k} \frac{1}{(k^{2}+m^{2})^{2}}\ ,
\end{equation}
and is calculated as follows:
\begin{eqnarray}\label{lf14}
I_{1} &=&\int_{k} \int_{0}^{\infty } \rmd \alpha \, \alpha \,\rme^{{-\alpha
(k^{2}+m^{2})} }\nn \\
 &=& \left(\int_{k}\rme^{-k^2} \right)  \int_{0}^{\infty
} \rmd \alpha\,  \alpha^{{1-\frac{d}{2}}} \, \rme^{-\alpha m^{2}}\nn \\
&=&  \left(\int_{k}\rme^{-k^2} \right) m^{-\epsilon } \Gamma
\left(\frac{\epsilon }{2} \right)\ .
\end{eqnarray}
We will also denote the dimensionless integral \begin{equation}
\tilde I_1 = I_1\Big|_{m=1}
\ .\end{equation}
This gives us the normalization-constant for higher-loop calculations
\begin{equation}
\left(\E I_1 \right) = m^{-\E}  \left(\int_{k}\rme^{-k^2} \right) \E
\Gamma \left(\frac{\epsilon }{2} \right) = m^{-\E}  \left(\int_{k}\rme^{-k^2} \right) 2\,
\Gamma \left(1+\frac{\epsilon }{2} \right)\ .
\end{equation}

\section{2-loop integral for the 2-point correlation function\label{app:TwoPointCorrelation}}
\begin{figure}
\begin{center}
\includegraphics[width=0.5\textwidth,angle=0]{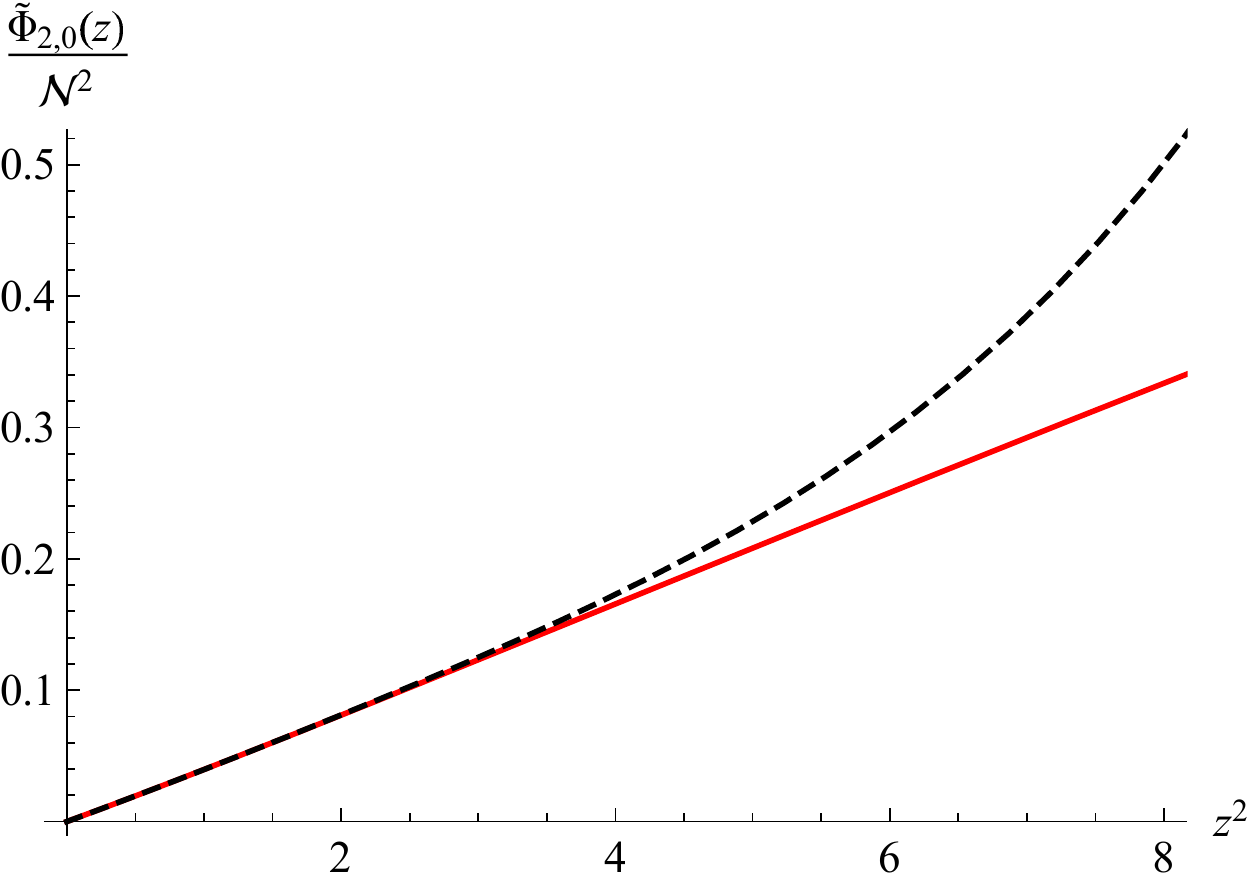}
\caption{Taylor expansion to 8th order as given in Eq.~(\ref{D:5})--(\ref{D:8}) (dashed curve) of  $\tilde{\Phi}_{2,0}(z)$ (red, solid line).}\label{fig:twoloopCorrTaylor}
\end{center}
\end{figure}

We consider the following 2-loop contribution to the 2-point correlation function
\begin{align}
 \Phi_{2,\epsilon}(q)&=I_A(q)-I_A + I_1^2 - I_1I_1(q) =  \sfrac{1}{m^{2\epsilon}}\tilde{\Phi}_{2,
\epsilon}(\sfrac{q}{m}) \, ,
\end{align}
which can be written as
\begin{align}
\tilde\Phi_{2,\epsilon}(z)&= { \ca N^2}\big[ F_{\epsilon}(z,1)-F_{\epsilon}(0,1)-F_{\epsilon}(z,0)+F_{\epsilon}(0,0)\big]\ ,
\end{align}
with
\begin{equation}
F_{\epsilon}(z,b)= \Gamma(\epsilon) \int_{x_1,x_2,x_3>0}  \frac{\left[1+x_1+x_2+x_3+
z^2\frac{x_1
   \left(x_2+x_3\right) + x_2 x_3 b^2}{(x_2+x_3)(1+x_1) +x_2x_3b^2}\right]^{{ -\epsilon}}
}{\Big[(x_2+x_3)(1+x_1) +x_2x_3b^2 \Big]^{2-\frac{\epsilon}{2}} }\ .
\end{equation}
Although each individual term is of order $\frac{1}{\epsilon^2}$, the limit $\epsilon\to 0$ of $\tilde{\Phi}_{2,\epsilon}$ exists and is given by
\begin{align}\label{eq:twoloopcorrPhi}
\tilde{\Phi}_{2,0}(z) &={\ca N^2} \int_{x_1,x_2,x_3>0}\left\{ \sfrac{ \ln \left(1+\frac{ z^2x_1}{(1+x_1)(1+x_1+x_2+x_3)}\right)}{(1+x_1)^2(x_2+x_3)^2} -  \sfrac{\ln\left(1+\frac{ z^2 x_1(x_2+x_3)+x_2x_3}{\big[(1+x_1)(x_2+x_3)+x_2x_3\big](1+x_1+x_2+x_3)} \right)}{\big[(x_2+x_3)(1+x_1) +x_2x_3\big]^2} \right\}\, .
\end{align}
We were not able to obtain a closed analytical expression for the three-dimensional integral. Using the variable transformations $x_1=\sfrac{1}{x}-1$, $x_3=\sfrac{y}{x}$, and $x_2=\sfrac{y_2}{x}$ helps to determine the Taylor expansion $ \tilde{\Phi}_{2,0}(z) \approx { \ca N^2} \sum_n \alpha_n z^{2n}$ with the first four coefficients
\begin{align}
\label{D:5}
 \alpha_1 &= -\frac 29 -\frac{8\pi^2}{243} + \frac{1}{81} \left[\psi'(\sfrac{1}{3}) + \psi'(\sfrac{1}{6})\right] \approx  0.03821\ ,\\
\alpha_2 &= \frac{193}{3240} +\frac{16\pi^2}{2187} - \frac{2}{729} \left[\psi'(\sfrac{1}{3}) + \psi'(\sfrac{1}{6})\right] \approx 0.00169\ ,\\
\alpha_3& \approx -0.00039\ ,
 \\ \alpha_4&\approx 0.00007 \ .
\label{D:8}
\end{align}
 Since the coefficients are small, a Taylor expansion to fourth order compares well with the full function up to $z\approx 3$, see Fig. \ref{fig:twoloopCorrTaylor}.

\begin{figure}
\begin{center}
\includegraphics[width=0.6\textwidth,angle=0]{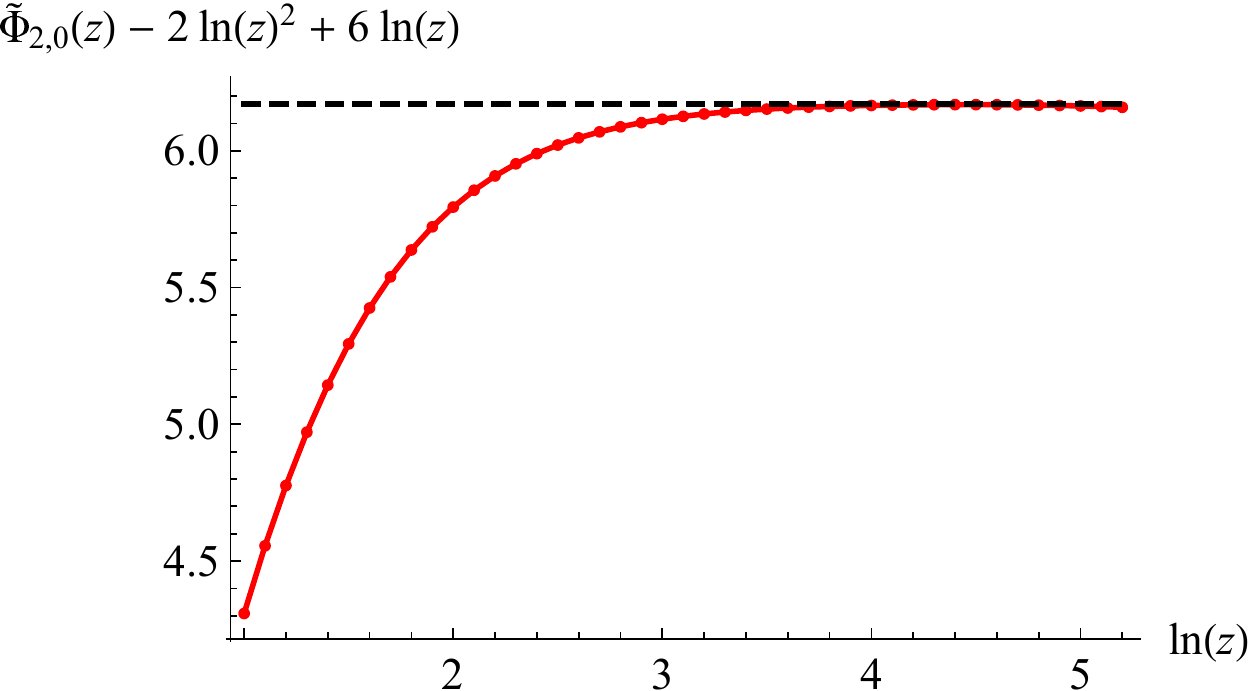}
\caption{Asymptotics of $\tilde{\Phi}_{2,0}(z) - 2 \ln (z)^2 +6\ln (z)$ for large $z$. The dashed line is the asypmptotic value $6.17$.}
\label{fig:twoloopCorrasymp}
\end{center}
\end{figure}

To render the integral numerically well-behaved, it is  convenient to perform a variable transformation to $s=x_2+x_3$, $d s=x_2-x_3$, \(x=x_1\). The integral to be calculated then is
(using the symmetry $d\to -d$)
\begin{equation}\tilde{\Phi}_{2,0}(z) = { \ca N^2}\int_{0<d<1}\int_{s,x>0} \frac{\ln\! \left(\frac{ x z^2}{\left(x+1\right)
   \left(s+x+1\right)}+1\right)}{s
   \left(x+1\right){}^2}-\frac{16 \ln \!\left(\frac{ z^2
   \left[\left(d^2-1\right) s-4
   x\right]}{\left(s+x+1\right) \left[\left(d^2-1\right)
   s-4 \left(x+1\right)\right]}+1\right)}{s \left(d^2 s-s-4
   x-4\right)^2}\ .
\end{equation}
In order to calculate the asymptotics for large $z$ we consider $z^2\sfrac{\mathrm{d}}{\mathrm{d}z^2} \tilde{\Phi}_{2,0}(z)$, which helps to solve the integrals but eliminates the constant part. Using again the variable transformations  $x_1=\sfrac{1}{x}-1$, $x_3=\sfrac{y}{x}$, and $x_2=\sfrac{y_2}{x}$ the integral reads \begin{align}
 z^2\sfrac{\mathrm{d}}{\mathrm{d}(z^2)} \tilde{\Phi}_{2,0}(z) &= {\ca N^2}\int_{0}^{1} \mathrm{d}x\int_{0}^{\infty} \mathrm{d}y \int_{0}^{\infty} \mathrm{d}y_2 \;\big[ F_z^{(1)}(x,y,y_2)+F_z^{(2)}(x,y,y_2) \big]\\
F_z^{(1)}(x,y,y_2) &=  \frac{x z^2(1 - x)}{(y + y_2)^2 (1 + y + y_2 + (1 - x) x z^2)}\\
F_z^{(2)}(x,y,y_2) &=  \frac{-x z^2[y (1 - x + y_2) + (1 - x) y_2]}{(y + y_2 +
   y y_2)^2 \big[(1 + y + y_2) (y + y_2 + y y_2) +
   xz ^2(y_2 - x y_2 + y (1 - x + y_2)) \big]}
\end{align}
We distinguish the cases $y<1$ and $y>1$ and split $z^2\sfrac{\mathrm{d}}{\mathrm{d}z^2} \tilde{\Phi}_{2,0}(z)=A_<+A_>$ accordingly. For $y<1$ the limit $z\to \infty$ exists and can be taken in the integrand. This integration gives   a constant,
\begin{align}
 \lim_{z\to \infty} A_< =  \int_{0}^{1} \mathrm{d}x\int_{0}^{\infty} \mathrm{d}y \int_{0}^{1} \mathrm{d}y_2 \;\left[\frac{1}{(y + y_2)^2} - \frac{1}{(y + y_2 + y y_2)^2}\right]= \ln 2\ .
\end{align}
  For $y>1$ we perform the $y$ and $y_2$ integration over the first term in Eq. (\ref{eq:twoloopcorrPhi}), then expand to lowest orders in $\frac 1z$ and integrate over $x$. This gives the logarithm
\begin{align}
 \int_{0}^{1} \mathrm{d}x\int_{0}^{\infty} \mathrm{d}y \int_{0}^{\infty}\mathrm{d}y_2 \;F_z^{(1)}(x,y,y_2) &= -3 + 2\ln z +\ca{O}(z\ln z).
\end{align}
Obtaining the next order is more delicate than expanding in $\frac 1z$ before the $x$-integration.
The second term gives again only a constant,   and the limit $z\to\infty$ can be taken in the integrand
\begin{align}
 \lim_{z\to \infty}\int_{0}^{1} \mathrm{d}x\int_{0}^{\infty} \mathrm{d}y \int_{0}^{\infty} \mathrm{d}y_2\; F_z^{(2)}(x,y,y_2)= -\ln 2\ .
\end{align}
In summary, we find\begin{align}
z^2\frac \rmd{\rmd z^2}\tilde{\Phi}_{2,0}( z) &= {\ca N^2} \big[ 2\ln z -3+\ca{O}(\sfrac{1}{z}\ln
z) \big] \ .
\end{align}
And consequently after integration
\begin{align}
\tilde{\Phi}_{2,0}( z) &={ \ca N^2} \big[  2  (\ln z)^2 -6\ln z +\alpha_0 +\ca{O}(\sfrac{1}{z}\ln
z)\big]  .
\end{align}
We plot the asymptotics in Fig. \ref{fig:twoloopCorrasymp} and find numerically $\alpha_0\approx 6.17(2)$.

\newcommand{\doi}[2]{\href{http://dx.doi.org/#1}{#2}}
\newcommand{\arxiv}[1]{\href{http://arxiv.org/abs/#1}{#1}}

%

\begin{thebibliography}{10}

\bibitem{NarayanDSFisher1993a}
O.~Narayan and D.S. Fisher,
\newblock {\em Threshold critical dynamics of driven interfaces in random
  media},
\newblock \doi{10.1103/PhysRevB.48.7030}{\rm Phys. Rev. B {\bf 48} (1993)
  7030--42}\null.

\bibitem{NarayanDSFisher1992b}
O.~Narayan and D.S. Fisher,
\newblock {\em Critical behavior of sliding charge-density waves in 4-epsilon
  dimensions},
\newblock \doi{10.1103/PhysRevB.46.11520}{\rm Phys. Rev. B {\bf 46} (1992)
  11520--49}\null.

\bibitem{NarayanDSFisher1992a}
O.~Narayan and D.S. Fisher,
\newblock {\em Dynamics of sliding charge-density waves in 4-epsilon
  dimensions},
\newblock \doi{10.1103/PhysRevLett.68.3615}{\rm Phys. Rev. Lett. {\bf 68}
  (1992)   3615--18}\null.

\bibitem{LeschhornNattermannStepanowTang1997}
H.~Leschhorn, T.~Nattermann, S.~Stepanow  and L.-H. Tang,
\newblock {\em Driven interface depinning in a disordered medium},
\newblock \doi{10.1002/andp.19975090102}{\rm Annalen der Physik {\bf 509}
  (1997)   1--34}\null,
\newblock \arxiv{arXiv:cond-mat/9603114}.

\bibitem{NattermannStepanowTangLeschhorn1992}
T.~Nattermann, S.~Stepanow, L.-H. Tang  and H.~Leschhorn,
\newblock {\em Dynamics of interface depinning in a disordered medium},
\newblock \doi{10.1051/jp2:1992214}{\rm J. Phys. II (France) {\bf 2} (1992)
  1483--8}\null.

\bibitem{LeDoussalWieseChauve2003}
P.~Le Doussal, K.J. Wiese  and P.~Chauve,
\newblock {\em Functional renormalization group and the field theory of
  disordered elastic systems},
\newblock \doi{10.1103/PhysRevE.69.026112}{\rm Phys. Rev. E {\bf 69} (2004)
  026112}\null,
\newblock \arxiv{cond-mat/0304614}.

\bibitem{LeDoussalWieseChauve2002}
P.~Le Doussal, K.J. Wiese  and P.~Chauve,
\newblock {\em 2-loop functional renormalization group analysis of the
  depinning transition},
\newblock \doi{10.1103/PhysRevB.66.174201}{\rm Phys. Rev. B {\bf 66} (2002)
  174201}\null,
\newblock \arxiv{cond-mat/0205108}.

\bibitem{ChauveLeDoussalWiese2000a}
P.~Chauve, P.~Le Doussal  and K.J. Wiese,
\newblock {\em Renormalization of pinned elastic systems: How does it work
  beyond one loop?},
\newblock \doi{10.1103/PhysRevLett.86.1785}{\rm Phys. Rev. Lett. {\bf 86}
  (2001)   1785--1788}\null,
\newblock \arxiv{cond-mat/0006056}.

\bibitem{MiddletonLeDoussalWiese2006}
A.A. Middleton, P.~{Le~Doussal}  and K.J. Wiese,
\newblock {\em Measuring functional renormalization group fixed-point functions
  for pinned manifolds},
\newblock \doi{10.1103/PhysRevLett.98.155701}{\rm Phys. Rev. Lett. {\bf 98}
  (2007)   155701}\null,
\newblock \arxiv{cond-mat/0606160}.

\bibitem{WieseHusemannLeDoussal2018}
K.J. Wiese, C.~Husemann  and P.~{Le Doussal},
\newblock {\em Field theory of disordered elastic interfaces at 3-loop order:
  The $\beta$-function},
\newblock  \doi{10.1016/j.nuclphysb.2018.04.013}{Nucl. Phys. B 932 (2018) 540-588},
\newblock \arxiv{arXiv:1801.08483}.

\bibitem{Johansson2000}
K.~Johansson,
\newblock {\em Shape fluctuations and random matrices},
\newblock \doi{10.1007/s002200050027}{\rm Communications in Mathematical
  Physics {\bf 209} (2000)   437--76}\null,
\newblock \arxiv{math/9903134}.

\bibitem{PraehoferSpohn2000a}
M.~Pr{\"a}hofer and H.~Spohn,
\newblock {\em Universal distributions for growth processes in 1 + 1 dimensions
  and random matrices},
\newblock \doi{10.1103/PhysRevLett.84.4882}{\rm Phys. Rev. Lett. {\bf 84}
  (2000)   4882--4885}\null,
\newblock \arxiv{cond-mat/9912264}.

\bibitem{PraehoferSpohn2000}
M.~Pr{\"a}hofer and H.~Spohn,
\newblock {\em Statistical self-similarity of one-dimensional growth
  processes},
\newblock \doi{10.1016/S0378-4371(99)00517-8}{\rm Physica A {\bf 279} (2000)
  342--52}\null,
\newblock \arxiv{cond-mat/9910273}.

\bibitem{KPZ}
M.~Kardar, G.~Parisi  and Y.-C. Zhang,
\newblock {\em Dynamic scaling of growing interfaces},
\newblock \doi{10.1103/PhysRevLett.56.889}{\rm Phys. Rev. Lett. {\bf 56} (1986)
    889--892}\null.

\bibitem{FreyTaeuber1994}
E.~Frey and U.C. T{\"a}uber,
\newblock {\em Two-loop renormalization group analysis of the
  {Burgers-Kardar-Parisi-Zhang} equation},
\newblock \doi{10.1103/PhysRevE.50.1024}{\rm Phys. Rev. E {\bf 50} (1994)
  1024--1045}\null.

\bibitem{Laessig1995}
M.~L{\"a}ssig,
\newblock {\em On the renormalization of the {Kardar-Parisi-Zhang} equation},
\newblock \doi{10.1016/0550-3213(95)00268-W}{\rm Nucl. Phys. {\bf B 448} (1995)
    559--574}\null,
\newblock \arxiv{cond-mat/9501094}.

\bibitem{FreyTaeuberHwa1996}
E.~Frey, U.C. T{\"a}uber  and T.~Hwa,
\newblock {\em Mode coupling and renormalization group results for the noisy
  {Burgers} equation},
\newblock \doi{10.1103/PhysRevE.53.4424}{\rm Phys. Rev. E {\bf 53} (1996)
  4424}\null.

\bibitem{Wiese1997c}
K.J. Wiese,
\newblock {\em Critical discussion of the 2-loop calculations for the
  {KPZ}-equation},
\newblock \doi{10.1103/PhysRevE.56.5013}{\rm Phys. Rev. {\bf E 56} (1997)
  5013--5017}\null,
\newblock \arxiv{cond-mat/9706009}.

\bibitem{Wiese1998a}
K.J. Wiese,
\newblock {\em On the perturbation expansion of the {KPZ}-equation},
\newblock \doi{10.1023/B:JOSS.0000026730.76868.c4}{\rm J. Stat. Phys. {\bf 93}
  (1998)   143--154}\null,
\newblock \arxiv{cond-mat/9802068}.

\bibitem{MarinariPagnaniParisi2000}
E.~Marinari, A.~Pagnani  and G.~Parisi,
\newblock {\em Critical exponents of the {KPZ} equation via multi-surface
  coding numerical simulations},
\newblock \doi{10.1088/0305-4470/33/46/303}{\rm J. Phys. A {\bf 33} (2000)
  8181--92}\null.

\bibitem{PraehoferSpohn1997}
M.~Pr{\"a}hofer and H.~Spohn,
\newblock {\em An exactly solved model of three-dimensional surface growth in
  the anisotropic {KPZ} regime},
\newblock \doi{10.1007/BF02732423}{\rm J. Stat. Phys. {\bf 88} (1997)
  999--1012}\null,
\newblock \arxiv{cond-mat/9612209}.

\bibitem{Krug1997}
J.~Krug,
\newblock {\em Origins of scale invariance in growth processes},
\newblock \doi{10.1080/00018739700101498}{\rm Advances in Physics {\bf 46}
  (1997)   139--282}\null.

\bibitem{Mezard1997}
M.~Mezard,
\newblock {\em Disordered systems and {Burger's} turbulence},
\newblock \doi{10.1051/jp4:1998603}{\rm J. Phys. IV (France) {\bf 8} (1997)
  27--38}\null,
\newblock \arxiv{cond-mat/9801029}.

\bibitem{MedinaHwaKardarZhang1989}
E.~Medina, T.~Hwa, M.~Kardar  and Y.C. Zhang,
\newblock {\em Burgers equation with correlated noise: Renormalization-group
  analysis and applications to directed polymers and interface growth},
\newblock \doi{10.1103/PhysRevA.39.3053}{\rm Phys. Rev. {\bf A 39} (1989)
  3053}\null.

\bibitem{HwaFisher1994b}
T.~Hwa and D.S. Fisher,
\newblock {\em Anomalous fluctuations of directed polymers in random media},
\newblock \doi{10.1103/PhysRevB.49.3136}{\rm Phys. Rev. B {\bf 49} (1994)
  3136--54}\null,
\newblock \arxiv{cond-mat/9309016}.

\bibitem{BundschuhHwa2000}
R.~Bundschuh and T.~Hwa,
\newblock {\em An analytic study of the phase transition line in local sequence
  alignment with gaps},
\newblock \doi{10.1016/S0166-218X(00)00188-8}{\rm Discrete Applied Mathematics
  {\bf 104} (2000)   113--42}\null.

\bibitem{BundschuhHwa1999}
R.~Bundschuh and T.~Hwa,
\newblock {\em {RNA} secondary structure formation: a solvable model of
  heteropolymer folding},
\newblock \doi{10.1103/PhysRevLett.83.1479}{\rm Phys. Rev. Lett. {\bf 83}
  (1999)   1479--82}\null,
\newblock \arxiv{cond-mat/9903089}.

\bibitem{HwaLaessig1998}
T.~Hwa and M.~L{\"a}ssig,
\newblock {\em Optimal detection of sequence similarity by local alignment}.
\newblock RECOMB 98, pages 109--16, 1998,
\newblock \arxiv{cond-mat/9712081}.

\bibitem{NattermannBookYoung}
T.~Nattermann,
\newblock {\em Theory of the random field {Ising} model},
\newblock in A.P. Young, editor, {\em Spin glasses and random fields}, World
  Scientific, Singapore, 1997,
\newblock \arxiv{cond-mat/9705295}.

\bibitem{LemerleFerreChappertMathetGiamarchiLeDoussal1998}
S.~Lemerle, J.~{Ferr\'e}, C.~Chappert, V.~Mathet, T.~Giamarchi  and P.~{Le
  Doussal},
\newblock {\em Domain wall creep in an {Ising} ultrathin magnetic film},
\newblock \doi{10.1103/PhysRevLett.80.849}{\rm Phys. Rev. Lett. {\bf 80} (1998)
    849}\null.

\bibitem{Gruner1988}
G.~Gr\"uner,
\newblock {\em The dynamics of charge-density waves},
\newblock \doi{10.1103/RevModPhys.60.1129}{\rm Rev. Mod. Phys. {\bf 60} (1988)
   1129--81}\null.

\bibitem{BlatterFeigelmanGeshkenbeinLarkinVinokur1994}
G.~Blatter, M.V. {Feigel'man}, V.B. Geshkenbein, A.I. Larkin  and V.M. Vinokur,
\newblock {\em Vortices in high-temperature superconductors},
\newblock \doi{10.1103/RevModPhys.66.1125}{\rm Rev. Mod. Phys. {\bf 66} (1994)
   1125}\null.

\bibitem{GiamarchiLeDoussalBookYoung}
T.~Giamarchi and P.~{Le~Doussal},
\newblock {\em Statics and dynamics of disordered elastic systems},
\newblock in A.P. Young, editor, {\em Spin glasses and random fields}, World
  Scientific, Singapore, 1997,
\newblock \arxiv{cond-mat/9705096}.

\bibitem{GiamarchiLeDoussal1995}
T.~Giamarchi and P.~Le Doussal,
\newblock {\em Elastic theory of flux lattices in the presence of weak
  disorder},
\newblock \doi{10.1103/PhysRevB.52.1242}{\rm Phys. Rev. B {\bf 52} (1995)
  1242--70}\null,
\newblock \arxiv{cond-mat/9501087}.

\bibitem{GiamarchiLeDoussal1994}
T.~Giamarchi and P.~Le Doussal,
\newblock {\em Elastic theory of pinned flux lattices},
\newblock \doi{10.1103/PhysRevLett.72.1530}{\rm Phys. Rev. Lett. {\bf 72}
  (1994)   1530--3}\null.

\bibitem{NattermannScheidl2000}
T.~Nattermann and S.~Scheidl,
\newblock {\em Vortex-glass phases in type-{II} superconductors},
\newblock \doi{10.1080/000187300412257}{\rm Advances in Physics {\bf 49} (2000)
    607--704}\null,
\newblock \arxiv{cond-mat/0003052}.

\bibitem{LeDoussalWieseMoulinetRolley2009}
P.~Le Doussal, K.J. Wiese, S.~Moulinet  and E.~Rolley,
\newblock {\em Height fluctuations of a contact line: {A} direct measurement of
  the renormalized disorder correlator},
\newblock \doi{10.1209/0295-5075/87/56001}{\rm EPL {\bf 87} (2009)
  56001}\null,
\newblock \arxiv{arXiv:0904.4156}.

\bibitem{PrevostRolleyGuthmann2002}
A.~Prevost, E.~Rolley  and C.~Guthmann,
\newblock {\em Dynamics of a helium-4 meniscus on a strongly disordered cesium
  substrate},
\newblock \doi{10.1103/PhysRevB.65.064517}{\rm Phys. Rev. B {\bf 65} (2002)
  064517/1--8}\null.

\bibitem{PrevostThese}
A.~Prevost,
\newblock PhD thesis, Orsay, 1999.

\bibitem{ErtasKardar1994b}
D.~Ertas and M.~Kardar,
\newblock {\em Critical dynamics of contact line depinning},
\newblock \doi{10.1103/PhysRevE.49.R2532}{\rm Phys. Rev. E {\bf 49} (1994)
  2532}\null.

\bibitem{LeDoussalWiese2009a}
P.~{Le Doussal} and K.J. Wiese,
\newblock {\em Elasticity of a contact-line and avalanche-size distribution at
  depinning},
\newblock \doi{10.1103/PhysRevE.82.011108}{\rm Phys. Rev. E {\bf 82} (2010)
  011108}\null,
\newblock \arxiv{arXiv:0908.4001}.

\bibitem{KardarLH1994}
M.~Kardar,
\newblock {\em Lectures on directed paths in random media},
\newblock in F.~David, P.~Ginsparg  and J.~Zinn-Justin, editors, {\em
  Fluctuating Geometries in Statistical Mechanics and Field Theory}, {\em {\em
  Volume} LXII} of {\em Les Houches, \'ecole d'\'et\'e de physique th\'eorique
  1994}, Elsevier Science, Amsterdam, 1996.

\bibitem{AragonKoltonDoussalWieseJagla2016}
L.E. Aragon, A.B. Kolton, P.~Le Doussal, K.J. Wiese  and E.~Jagla,
\newblock {\em Avalanches in tip-driven interfaces in random media},
\newblock \doi{10.1209/0295-5075/113/10002}{\rm EPL {\bf 113} (2016)
  10002}\null,
\newblock \arxiv{arXiv:1510.06795}.

\bibitem{DurinBohnCorreaSommerDoussalWiese2016}
G.~Durin, F.~Bohn, M.A. Correa, R.L. Sommer, P.~Le Doussal  and K.J. Wiese,
\newblock {\em Quantitative scaling of magnetic avalanches},
\newblock \doi{10.1103/PhysRevLett.117.087201}{\rm Phys. Rev. Lett. {\bf 117}
  (2016)   087201}\null,
\newblock \arxiv{arXiv:1601.01331}.

\bibitem{LaursonIllaSantucciTallakstadyAlava2013}
L.~Laurson, X.~Illa, S.~Santucci, K.T. Tallakstad, K.J.~M\aa l\o y  and M.J.
  Alava,
\newblock {\em Evolution of the average avalanche shape with the universality
  class},
\newblock \doi{10.1038/ncomms3927}{\rm Nat. Commun. {\bf 4} (2013)
  2927}\null.

\bibitem{DurinZapperi2000}
G.~Durin and S.~Zapperi,
\newblock {\em Scaling exponents for {Barkhausen} avalanches in polycrystalline
  and amorphous ferromagnets},
\newblock \doi{10.1103/PhysRevLett.84.4705}{\rm Phys. Rev. Lett. {\bf 84}
  (2000)   4705--4708}\null.

\bibitem{PerkovicDahmenSethna1995}
O.~Perkovic, K.~Dahmen  and JP. Sethna,
\newblock {\em Avalanches, {Barkhausen} noise, and plain old criticality},
\newblock \doi{10.1103/PhysRevLett.75.4528}{\rm Phys. Rev. Lett. {\bf 75}
  (1995)   4528--4531}\null.

\bibitem{LeDoussalWiese2012a}
P.~Le~Doussal and K.J. Wiese,
\newblock {\em Avalanche dynamics of elastic interfaces},
\newblock \doi{10.1103/PhysRevE.88.022106}{\rm Phys. Rev. E {\bf 88} (2013)
  022106}\null,
\newblock \arxiv{arXiv:1302.4316}.

\bibitem{DobrinevskiLeDoussalWiese2013}
A.~Dobrinevski, P.~Le~Doussal  and K.J. Wiese,
\newblock {\em Statistics of avalanches with relaxation and {Barkhausen} noise:
  A solvable model},
\newblock \doi{10.1103/PhysRevE.88.032106}{\rm Phys. Rev. E {\bf 88} (2013)
  032106}\null,
\newblock \arxiv{arXiv:1304.7219}.

\bibitem{DobrinevskiLeDoussalWiese2011b}
A.~Dobrinevski, P.~{Le Doussal}  and K.J. Wiese,
\newblock {\em Non-stationary dynamics of the
  {Alessandro-Beatrice-Bertotti-Montorsi} model},
\newblock \doi{10.1103/PhysRevE.85.031105}{\rm Phys. Rev. E {\bf 85} (2012)
  031105}\null,
\newblock \arxiv{arXiv:1112.6307}.

\bibitem{LeDoussalMuellerWiese2011}
P.~Le Doussal, M.~M\"uller  and K.J. Wiese,
\newblock {\em Equilibrium avalanches in spin glasses},
\newblock \doi{10.1103/PhysRevB.85.214402}{\rm Phys. Rev. B {\bf 85} (2012)
  214402}\null,
\newblock \arxiv{arXiv:1110.2011}.

\bibitem{LeDoussalWiese2008a}
P.~Le Doussal and K.J. Wiese,
\newblock {\em Driven particle in a random landscape: disorder correlator,
  avalanche distribution and extreme value statistics of records},
\newblock \doi{10.1103/PhysRevE.79.051105}{\rm Phys. Rev. E {\bf 79} (2009)
  051105}\null,
\newblock \arxiv{arXiv:0808.3217}.

\bibitem{ZhuWiese2017}
Z.~Zhu and {K.J.} Wiese,
\newblock {\em The spatial shape of avalanches},
\newblock \doi{10.1103/PhysRevE.96.062116}{\rm Phys. Rev. E {\bf 96} (2017)
  062116}\null,
\newblock \arxiv{arXiv:1708.01078}.

\bibitem{WieseFedorenko2018}
{K.J.} Wiese and A.A. Fedorenko,
\newblock {\em Field theories for loop-erased random walks},
\newblock (2018),
\newblock \arxiv{arXiv:1802.08830}.

\bibitem{LeDoussal2006b}
P.~{Le Doussal},
\newblock {\em Finite temperature {Functional RG}, droplets and decaying
  {Burgers} turbulence},
\newblock \doi{10.1209/epl/i2006-10295-1}{\rm Europhys. Lett. {\bf 76} (2006)
  457--463}\null,
\newblock \arxiv{cond-mat/0605490}.

\bibitem{LeDoussalWiese2006a}
P.~{Le Doussal} and K.J. Wiese,
\newblock {\em How to measure {Functional RG} fixed-point functions for
  dynamics and at depinning},
\newblock \doi{10.1209/0295-5075/77/66001}{\rm EPL {\bf 77} (2007)
  66001}\null,
\newblock \arxiv{cond-mat/0610525}.

\bibitem{RossoLeDoussalWiese2006a}
A.~Rosso, P.~{Le~Doussal}  and K.J. Wiese,
\newblock {\em Numerical calculation of the functional renormalization group
  fixed-point functions at the depinning transition},
\newblock \doi{10.1103/PhysRevB.75.220201}{\rm Phys. Rev. B {\bf 75} (2007)
  220201}\null,
\newblock \arxiv{cond-mat/0610821}.

\bibitem{LeDoussalMuellerWiese2007}
P.~{Le Doussal}, M.~M\"uller  and K.J. Wiese,
\newblock {\em Cusps and shocks in the renormalized potential of glassy random
  manifolds: How functional renormalization group and replica symmetry breaking
  fit together},
\newblock \doi{10.1103/PhysRevB.77.064203}{\rm Phys. Rev. B {\bf 77} (2007)
  064203}\null,
\newblock \arxiv{arXiv:0711.3929}.

\bibitem{LeDoussalWiese2011b}
P.~{Le Doussal} and K.J. Wiese,
\newblock {\em First-principle derivation of static avalanche-size
  distribution},
\newblock \doi{10.1103/PhysRevE.85.061102}{\rm Phys. Rev. E {\bf 85} (2011)
  061102}\null,
\newblock \arxiv{arXiv:1111.3172}.

\bibitem{LeDoussalMiddletonWiese2008}
P.~{Le~Doussal}, A.A. Middleton  and K.J.\ Wiese,
\newblock {\em Statistics of static avalanches in a random pinning landscape},
\newblock \doi{10.1103/PhysRevE.79.050101}{\rm Phys. Rev. E {\bf 79} (2009)
  050101 (R)}\null,
\newblock \arxiv{arXiv:0803.1142}.

\bibitem{LeDoussalWiese2008c}
P.~{Le~Doussal} and K.J. Wiese,
\newblock {\em Size distributions of shocks and static avalanches from the
  functional renormalization group},
\newblock \doi{10.1103/PhysRevE.79.051106}{\rm Phys. Rev. E {\bf 79} (2009)
  051106}\null,
\newblock \arxiv{arXiv:0812.1893}.

\bibitem{RossoLeDoussalWiese2009a}
A.~Rosso, P.~{Le~Doussal}  and K.J.\ Wiese,
\newblock {\em Avalanche-size distribution at the depinning transition: A
  numerical test of the theory},
\newblock \doi{10.1103/PhysRevB.80.144204}{\rm Phys. Rev. B {\bf 80} (2009)
  144204}\null,
\newblock \arxiv{arXiv:0904.1123}.

\bibitem{KardarHuseHenleyFisher1985}
M.~Kardar, D.A. Huse, C.L. Henley  and D.S. Fisher,
\newblock {\em Roughening by impurities at finite temperatures (comment and
  reply)},
\newblock \doi{10.1103/PhysRevLett.55.2923}{\rm Phys. Rev. Lett. {\bf 55}
  (1985)   2923--4}\null.

\bibitem{Middleton1995}
A.A. Middleton,
\newblock {\em Numerical results for the ground-state interface in a random
  medium},
\newblock \doi{10.1103/PhysRevE.52.R3337}{\rm Phys. Rev. E {\bf 52} (1995)
  R3337--40}\null.

\bibitem{LeDoussalMonthus2003}
P.~Le Doussal and C.~Monthus,
\newblock {\em Exact solutions for the statistics of extrema of some random 1d
  landscapes, application to the equilibrium and the dynamics of the toy
  model},
\newblock \doi{10.1016/S0378-4371(02)01317-1}{\rm Physica A {\bf 317} (2003)
  140--98}\null,
\newblock \arxiv{cond-mat/0204168}.

\bibitem{LeDoussalWiese2003a}
P.~Le Doussal and K.J. Wiese,
\newblock {\em Higher correlations, universal distributions and finite size
  scaling in the field theory of depinning},
\newblock \doi{10.1103/PhysRevE.68.046118}{\rm Phys. Rev. E {\bf 68} (2003)
  046118}\null,
\newblock \arxiv{cond-mat/0301465}.

\bibitem{GrassbergerDharMohanty2016}
P.~Grassberger, D.~Dhar  and P.~K. Mohanty,
\newblock {\em Oslo model, hyperuniformity, and the quenched
  {Edwards-Wilkinson} model},
\newblock \doi{10.1103/PhysRevE.94.042314}{\rm Phys. Rev. E {\bf 94} (2016)
  042314}\null.

\bibitem{FerreroBustingorryKolton2012}
E.E. Ferrero, S. Bustingorry  and A.B. Kolton,
\newblock {\em Non-steady relaxation and critical exponents at the depinning
  transition},
\newblock \doi{10.1103/PhysRevE.87.032122}{\rm Phys. Rev. E {\bf 87} (2013)
  032122}\null,
\newblock \arxiv{arXiv:1211.7275}.

\bibitem{LeDoussalWiese2001}
P.~Le Doussal and K.J. Wiese,
\newblock {\em Functional renormalization group at large {$N$} for random
  manifolds},
\newblock \doi{10.1103/PhysRevLett.89.125702}{\rm Phys. Rev. Lett. {\bf 89}
  (2002)   125702}\null,
\newblock \arxiv{cond-mat/0109204}.

\bibitem{LeDoussalWiese2003b}
P.~Le Doussal and K.J. Wiese,
\newblock {\em Functional renormalization group at large ${N}$ for disordered
  elastic systems, and relation to replica symmetry breaking},
\newblock \doi{10.1103/PhysRevB.68.174202}{\rm Phys. Rev. B {\bf 68} (2003)
  174202}\null,
\newblock \arxiv{cond-mat/0305634}.

\bibitem{LeDoussalWiese2004a}
P.~{Le Doussal} and K.J. Wiese,
\newblock {\em Derivation of the functional renormalization group
  $\beta$-function at order $1/{N}$ for manifolds pinned by disorder},
\newblock \doi{10.1016/j.nuclphysb.2004.08.022}{\rm Nucl. Phys. B {\bf 701}
  (2004)   409--480}\null,
\newblock \arxiv{cond-mat/0406297}.

\end{thebibliography}
%

\tableofcontents

\end{document}